\documentclass[fleqn,usenatbib]{mnras}
\usepackage{newtxtext,newtxmath}
\usepackage{siunitx}
\usepackage{amsmath}
\usepackage{appendix}
\usepackage{comment}
\usepackage[T1]{fontenc}
\usepackage{ae,aecompl}
\usepackage{comment}
\usepackage[dvipsnames]{xcolor}

\usepackage{graphicx}	
\usepackage{amsmath}	
\usepackage{multirow}
\usepackage{color}
\usepackage[dvipsnames]{xcolor}
\usepackage{threeparttable, tablefootnote}
\usepackage{tabularx}
\usepackage{isotope}
\usepackage{subfigure}
\usepackage{lineno}
\usepackage{lastpage}
\usepackage{eso-pic}

\usepackage{etoolbox}
\def\reff@jnl#1{{\rm#1\/}}
\def\aj{\reff@jnl{AJ}}                 
\def\araa{\reff@jnl{ARA\&A}}           
\def\apj{\reff@jnl{ApJ}}               
\def\apjl{\reff@jnl{ApJ}}              
\def\apjs{\reff@jnl{ApJS}}             
\def\ao{\reff@jnl{Appl.Optics}}        
\def\apss{\reff@jnl{Ap\&SS}}           
\def\AAp{\reff@jnl{A\&A}}              
\def\AApr{\reff@jnl{A\&A~Rev.}}        
\def\AAps{\reff@jnl{A\&AS}}            
\def\azh{\reff@jnl{AZh}}               
\def\baas{\reff@jnl{BAAS}}             
\def\jcap{\reff@jnl{JCAP}}             
\def\jrasc{\reff@jnl{JRASC}}           
\def\mnras{\reff@jnl{MNRAS}}           
\def\pra{\reff@jnl{Phys.Rev.A}}        
\def\prb{\reff@jnl{Phys.Rev.B}}        
\def\prc{\reff@jnl{Phys.Rev.C}}        
\def\prd{\reff@jnl{Phys.Rev.D}}        
\def\prl{\reff@jnl{Phys.Rev.Lett}}     
\def\pasp{\reff@jnl{PASP}}             
\def\pasj{\reff@jnl{PASJ}}             
\def\qjras{\reff@jnl{QJRAS}}           
\def\skytel{\reff@jnl{S\&T}}           
\def\solphys{\reff@jnl{Solar~Phys.}}   
\def\sovast{\reff@jnl{Soviet~Ast.}}    
\def\ssr{\reff@jnl{Space~Sci.Rev.}}    
\def\zap{\reff@jnl{ZAp}}               
\def\nat{\reff@jnl{Nature}}            

\def\snana{\textsc{snana}}

\def\autoscan{\textsc{autoscan}}
\def\DiffImg{\textsc{diffimg}}
\def\baselineCC{6.6 }
\def\minCC{5.8 }
\def\maxCC{9.3 }
\def\avgCC{7.0 }
\def\rmsCC{1.1 }

\defcitealias{2014AJ....147...99M}{M14}
\defcitealias{2014ApJS..213...19B}{B14}
\defcitealias{2017ApJS..233....6H}{H17}
\defcitealias{2014Ap&SS.354...89B}{Br14}
\defcitealias{2011MNRAS.412.1441L}{L11}
\defcitealias{2014AJ....147..118R}{R14}
\defcitealias{Jones_2017_I}{J17}
\defcitealias{Hlozek_2012}{H12}
\defcitealias{Shivvers_2017}{S17}
\defcitealias{DES_biascor}{K19}
\defcitealias{Vincenzi_2019}{V19}

\newcommand{\rchisq}{\ensuremath{\chi^2_\nu}}
\newcommand{\effz}{$\varepsilon_{z_{\text{spec}}}$}

\title[Selection efficiency and core collapse supernova simulations]
{The Dark Energy Survey Supernova Program: Modelling selection efficiency and observed core collapse supernova contamination}

\AddToShipoutPictureBG*{%
  \AtPageUpperLeft{%
    \hspace{0.75\paperwidth}%
    \raisebox{-3.5\baselineskip}{%
      \makebox[0pt][l]{\textnormal{DES-2020-0522}}
}}}%

\AddToShipoutPictureBG*{%
  \AtPageUpperLeft{%
    \hspace{0.75\paperwidth}%
    \raisebox{-4.5\baselineskip}{%
      \makebox[0pt][l]{\textnormal{FERMILAB-PUB-20-629-AE}}
}}}%

\begin{document}
\label{firstpage}
\pagerange{\pageref{firstpage}--\pageref{lastpage}}

\author[Vincenzi et al.]{
\parbox{\textwidth}{
\Large
M.~Vincenzi,$^{1}$\footnote{email}
M.~Sullivan,$^{2}$
O.~Graur,$^{1,3}$
D.~Brout,$^{4,5}$
T.~M.~Davis,$^{6}$
C.~Frohmaier,$^{1}$
L.~Galbany,$^{7}$
C.~P.~Guti\'errez,$^{2}$
S.~R.~Hinton,$^{6}$
R.~Hounsell,$^{8}$
L.~Kelsey,$^{2}$
R.~Kessler,$^{9,10}$
E.~Kovacs,$^{11}$
S.~Kuhlmann,$^{11}$
J.~Lasker,$^{9,10}$
C.~Lidman,$^{12,13}$
A.~M\"oller,$^{14}$
R.~C.~Nichol,$^{1}$
M.~Sako,$^{15}$
D.~Scolnic,$^{16}$
M.~Smith,$^{2}$
E.~Swann,$^{1}$
P.~Wiseman,$^{2}$
J.~Asorey,$^{17}$
G.~F.~Lewis,$^{18}$
R.~Sharp,$^{13}$
B.~E.~Tucker,$^{13}$
M.~Aguena,$^{19,20}$
S.~Allam,$^{21}$
S.~Avila,$^{22}$
E.~Bertin,$^{23,24}$
D.~Brooks,$^{25}$
D.~L.~Burke,$^{26,27}$
A.~Carnero~Rosell,$^{28,29}$
M.~Carrasco~Kind,$^{30,31}$
J.~Carretero,$^{32}$
F.~J.~Castander,$^{33,34}$
A.~Choi,$^{35}$
M.~Costanzi,$^{36,37}$
L.~N.~da Costa,$^{20,38}$
M.~E.~S.~Pereira,$^{39}$
J.~De~Vicente,$^{17}$
S.~Desai,$^{40}$
H.~T.~Diehl,$^{21}$
P.~Doel,$^{25}$
S.~Everett,$^{41}$
I.~Ferrero,$^{42}$
P.~Fosalba,$^{33,34}$
J.~Frieman,$^{21,10}$
J.~Garc\'ia-Bellido,$^{22}$
E.~Gaztanaga,$^{33,34}$
D.~W.~Gerdes,$^{43,39}$
D.~Gruen,$^{44,26,27}$
R.~A.~Gruendl,$^{30,31}$
G.~Gutierrez,$^{21}$
D.~L.~Hollowood,$^{41}$
K.~Honscheid,$^{35,45}$
B.~Hoyle,$^{46,47,48}$
D.~J.~James,$^{49}$
K.~Kuehn,$^{50,51}$
N.~Kuropatkin,$^{21}$
M.~A.~G.~Maia,$^{20,38}$
P.~Martini,$^{35,52,53}$
F.~Menanteau,$^{30,31}$
R.~Miquel,$^{54,32}$
R.~Morgan,$^{55}$
A.~Palmese,$^{21,10}$
F.~Paz-Chinch\'{o}n,$^{56,31}$
A.~A.~Plazas,$^{57}$
A.~K.~Romer,$^{58}$
E.~Sanchez,$^{17}$
V.~Scarpine,$^{21}$
S.~Serrano,$^{33,34}$
I.~Sevilla-Noarbe,$^{17}$
M.~Soares-Santos,$^{39}$
E.~Suchyta,$^{59}$
G.~Tarle,$^{39}$
D.~Thomas,$^{1}$
C.~To,$^{44,26,27}$
T.~N.~Varga,$^{47,48}$
A.~R.~Walker,$^{60}$
and R.D.~Wilkinson$^{58}$
\begin{center} (DES Collaboration) \end{center}
}
}
\date{$\star$ maria.vincenzi@port.ac.uk\\ Author affiliations are shown in Appendix \ref{aff}\\
 Accepted XXX. Received YYY; in original form ZZZ\\}


\maketitle
\begin{abstract}

The analysis of current and future cosmological surveys of type Ia supernovae (SNe Ia) at high-redshift depends on the accurate photometric classification of the SN events detected.
Generating realistic simulations of photometric SN surveys constitutes an essential step for training and testing photometric classification algorithms, and for correcting biases introduced by selection effects and contamination arising from core collapse SNe in the photometric SN Ia samples. We use published SN time-series spectrophotometric templates, rates, luminosity functions and empirical relationships between SNe and their host galaxies to construct a framework for simulating photometric SN surveys. We present this framework in the context of the Dark Energy Survey (DES) 5-year photometric SN sample, comparing our simulations of DES with the observed DES transient populations. We demonstrate excellent agreement in many distributions, including Hubble residuals, between our simulations and data. 
We estimate the core collapse fraction expected in the DES SN sample after selection requirements are applied and before photometric classification. After testing different modelling choices and astrophysical assumptions underlying our simulation, we find that the predicted contamination varies from \minCC to \maxCC per cent, with an average of \avgCC per cent and r.m.s. of \rmsCC per cent.
Our simulations are the first to reproduce the observed photometric SN and host galaxy properties in high-redshift surveys without fine-tuning the input parameters. The simulation methods presented here will be a critical component of the cosmology analysis of the DES photometric SN Ia sample: correcting for biases arising from contamination, and evaluating the associated systematic uncertainty.

\end{abstract}


\section{Introduction}\label{sec:intro}

Type Ia supernovae (SNe~Ia) are a mature and well-understood cosmological probe via their use as standardisable candles \citep{scolnicnextgen}. They remain a uniquely powerful distance indicator in the high redshift universe, and directly constrain the properties of dark energy. When combined with \textit{Planck} cosmic microwave background (CMB) measurements, current SN Ia samples measure the dark energy equation-of-state parameter $w$ with a precision of $\sim0.05-0.06$ \citep{Betoule_2014,scolnic2018,DES_abbott_ref}, and show it to be consistent with a cosmological constant ($w\equiv-1$). 

With current and next generation SN surveys (DES, \citealp{2019PhRvL.122q1301A}; LSST, \citealp{2019ApJ...873..111I}; Nancy Grace Roman Space Telescope, formerly WFIRST, \citealp{2018ApJ...867...23H}), statistical uncertainties on SNe~Ia cosmological measurements are becoming comparable to systematic uncertainties \citep{DES_syst}.
In this paper, we tackle some of the most important sources of systematic uncertainty related to SN~Ia cosmological analysis and in particular we focus on core collapse contamination and selection effects. 

The Dark Energy Survey (DES) SN programme (DES SN) is the current state-of-the-art sample for SN Ia cosmology analysis. Over five seasons, this programme discovered and monitored more than 30,000 optical transients of various astrophysical origins.
For 60 per cent of this sample the spectroscopic redshift of the identified host galaxy has been measured \citep[many via the OzDES programme; see][]{lidman2020ozdes} and approximately 570 transients have been spectroscopically confirmed and classified \citep[e.g.,][]{DES_spec}. 

The first cosmological results using SNe~Ia from DES (DES-SN3YR) have been measured from a sample of 207 spectroscopically-confirmed SNe~Ia observed during the first three DES SN seasons, combined with 122 publicly available low-redshift SNe  \cite{DESCombined_2019, DES_abbott_ref, DES_h0}. Detailed descriptions of the analysis are presented by \cite{DES_SMP, DES_syst, DES_biascor, DES_chrom, DES_massstep}. The final 5-year DES SN sample will include not only spectroscopically-confirmed SNe~Ia, but also photometrically-identified SNe~Ia with a spectroscopic redshift measured from the identified host galaxy. This constitutes the DES photometric SN sample and it is an order of magnitude larger than the sample used for the first published cosmological results.
This increases the statistical power of the DES SN sample significantly, but with the complication of additional sources of systematic uncertainties that need to be considered, for example, those due to the photometric classification of the SNe, and due to the efficiency of measuring host galaxy redshifts.

The DES photometric SN sample includes a fraction of core collapse SN events photometrically similar to SNe~Ia but with a different astrophysical origin, and therefore different intrinsic brightnesses. Modelling this population of contaminants, and assessing the impact on cosmology, is one of the key challenges to fully exploit the DES photometric SN sample. This modelling is complex and depends on realistic simulations of core collapse SNe, which can be combined with simulations of SNe~Ia to build mock catalogues of the DES-SN sample. These simulations are used for modelling selection effects and biases, and to generate training samples for SN classification algorithms, i.e., algorithms designed to identify the type of a SN from photometric data alone. 

In the last decade, various SN photometric classifiers have been developed, and algorithms that exploit machine-learning techniques typically outperform other classifiers based on a template fitting approach  \citep[e.g.][]{Lochner_2016, Boone_2019, 2020MNRAS.491.4277M}.
However, the performance of machine-learning photometric classifiers is fundamentally dependent on homogeneous, representative and large training samples, with $>$100,000 events required in some cases. Unfortunately, spectroscopically confirmed SN samples are significantly more limited in size, usually biased towards brighter events and discovered in lower surface brightness local environments where it is easier to observe a spectrum with the signal-to-noise adequate for classification.

Using such spectroscopically-confirmed SN samples as training samples is therefore not a viable option, and instead representative training samples are typically generated with simulations.

For similar reasons, the validation and testing of photometric classifiers also requires realistic simulations and cannot be performed on data alone. However, the training, validation and testing of photometric classifiers on samples (either real or simulated) can lead to over-fitting and over-estimations of sample purity, particularly if the training samples contain only a limited snapshot of the true astrophysical diversity of the SN population.

Therefore, tests of the true performances of photometric classifiers must be carefully designed to avoid overestimating the accuracy of these algorithms and, for future cosmological analysis, this is ultimately as important as developing photometric classification algorithms. The methods presented here aim to address this critical validation issue.

There have been many attempts to improve the simulations of core collapse SNe. The initial set of core collapse templates published for the Supernova Photometric Classification Challenge \citep[SNPhotCC;][]{2010arXiv1001.5210K,2010PASP..122.1415K}, have been updated with models of type IIb SNe and SN1991bg-like SNe~Ia from \citet{Jones_2017_I} in order to augment the diversity of simulated contamination. The Photometric LSST Astronomical Time-Series Classification Challenge Team \citep[PLAsTiCC;][]{2018arXiv181000001T, 2019PASP..131i4501K} further improved and expanded this library, including other types of transients and exploring other techniques to augment template diversity. Independently, a new library of core collapse templates has been presented by \citet{Vincenzi_2019}. These templates are built from core collapse SNe using high-quality photometry and spectroscopy, and they have been robustly extended to ultraviolet (UV) wavelengths. Simulations also rely on core collapse SN luminosity functions and rates, for which several measurements have been recently published \citep[]{Strolger_2015,Shivvers_2017,Graur_2017, Vincenzi_2019, 2020arXiv201015270F}.

There are many elements of uncertainty in simulations of core collapse SNe, especially at intermediate and high redshift. Most measurements of core collapse SN demographics available in the literature are based on small and primarily low-redshift samples ($z\lesssim0.05$), whereas SN surveys like DES probe a significantly larger range in redshift ($z\lesssim1.2$).
For example, results from the Pan-STARRS Medium Deep Survey \citep{Jones_2017_I,Jones_2018_II} demonstrated that simulations based on currently published measurements of core collapse SN global properties, do not accurately reproduce the core collapse contamination observed in high-redshift Hubble residuals. They find that in order to reproduce the contamination observed in the Pan-STARRS photometric SN sample, the luminosity functions from \citet{2011MNRAS.412.1441L} need to be brightened by one magnitude, and the brightness dispersion for SNe Ib/c reduced by 55 per cent. 

Finally, the effects of inaccurate modelling of core collapse SNe are easily conflated with another important uncertainty in SN samples: selection effects. Simulations of photometric SN experiments like Pan-STARRS and DES require modelling of the SN detection efficiency and the efficiency of measuring host galaxy spectroscopic redshifts. While the SN detection efficiency has been robustly modelled for numerous surveys over the past decade using image-based simulations (e.g., \citealp{Dilday_2008, Perrett_2012}, and for DES, \citealp{2015AJ....150..172K}, \citealp{DES_biascor}), there is very limited work on how to model selection effects from host galaxy spectroscopic redshift surveys using a similar first principles modelling approach, and significant fine-tuning is usually applied.

In this paper, we present a set of realistic simulations of the DES photometric SN survey for which we significantly improve the modelling of core collapse SNe and of the efficiency of measuring spectroscopic redshifts of SN host galaxies. The improvements in the core collapse SN modelling are due to the implementation of high quality templates and other published measurements of global core collapse SN properties. To improve the modelling of the spectroscopic redshift efficiency, we explore a novel, data-driven approach and model the spectroscopic redshift efficiency as a function of host galaxy properties. We improve the simulation of SN host galaxies, and associate hosts to simulated SNe using published measurements of SN rates as a function of galaxy properties. 
The simulations presented in this paper constitute the foundation for a robust estimation of cosmological biases due to the core collapse SN contamination expected in the DES photometric SN sample.

We present an overview of the DES SN sample in Section~\ref{sec:data}, and describe how we estimate and model selection effects from the host spectroscopic redshift survey in Section~\ref{sec:specz_efficiency}. In Section~\ref{sec:simulations} we present the baseline approach to build simulations of the DES photometric SN sample. In Section~\ref{sec:data_vs_sims} we compare our simulations and the DES SN dataset and we evaluate how well our simulations reproduce core collapse SN contamination in the DES sample. In Section~\ref{sec:sim_cc_vary}, we test how sensitive our results are to our assumptions and the choices of template libraries used to generate core collapse SN simulations. We summarise in Section~\ref{sec:summary} and discuss future directions.


\section{The DES photometric SN sample}\label{sec:data}

DES is an optical imaging survey designed to constrain the properties of dark energy and other cosmological parameters by combining four different astrophysical probes: weak gravitational lensing, large scale structure, galaxy clusters and SNe~Ia \citep{2019PhRvL.122q1301A}. The imaging data are acquired by the Dark Energy Camera \citep[DECam;][]{2015AJ....150..150F}, mounted on the Blanco 4-m telescope at the Cerro Tololo Inter-American Observatory. DES surveyed 5000~deg$^2$ of the southern hemisphere sky over six years. For time-domain science, DES monitored ten 3-deg$^2$ fields with an average cadence of 7 days in the $griz$ filters during the first five years. Eight of these ten fields (X1, X2, E1, E2, C1, C2, S1, S2) were observed to a single-visit depth of $m\sim23.5$\,mag (\lq shallow fields\rq), and two (X3, C3) to a depth of $m\sim24.5$ mag (\lq deep fields\rq). 

In this section, we present the DES photometric SN sample. This is defined as the sample of SN Ia-like events discovered by DES over five years of observations and for which a spectroscopic redshift for the identified host has been obtained. The discovery and photometry of DES SNe are presented in Section~\ref{sec:discovery}, and the host galaxy identification and spectroscopic redshift measurements in Sections~\ref{sec:spec_followup} and \ref{sec:gal_ass}. In Section~\ref{sec:salt2_data}, we discuss how SN Ia-like events are selected from the data, and their light curves fitted using SN Ia spectra energy distribution (SED) models. In this analysis we neither discuss nor apply cuts based on SN~Ia photometric classifiers, which are often used in SN cosmological analysis to improve the purity of photometrically-selected SN samples. This is to intentionally enhance core collapse contamination in the DES sample and better analyse the properties of this population of contaminants.

\subsection{SN discovery and photometry}
\label{sec:discovery}
In DES-SN, the Difference Imaging pipeline \citep[\DiffImg,][]{2015AJ....150..172K} is used to discover and estimate the flux of new transients via image subtraction, comparing new observations with previously collected reference images. The detections are passed through an automated artefact rejection algorithm \citep[\autoscan;][]{2015AJ....150...82G}.

\DiffImg\ is an efficient tool for the rapid identification of transients and the estimation of their fluxes at the two per cent level. However, it does not provide photometric measurements at the level of precision and accuracy required for SN Ia cosmology. The DES-SN three-year (DES-SN3YR) cosmological analysis therefore used the technique of scene modelling photometry \citep[SMP;][]{Holtzman_2008, Astier_2013, DES_SMP}. The SMP algorithm simultaneously models the time-varying flux of a transient and the time-independent background flux from the host galaxy. SMP does not require image remapping and it determines robust uncertainties. However, it is computationally more expensive to run compared to \DiffImg.
The ongoing effort of running SMP on the full DES SN sample will be essential for cosmological measurements; however, \DiffImg\ photometry is adequate for developing many SNIa-cosmology analysis methods, including the methods presented in this paper.

We use as our initial sample of candidate SNe all DES events with at least two detections (in any filter, separated by at least one night) with a signal-to-noise ratio (SNR) greater than five, and that passed \autoscan. These criteria are designed to remove asteroids and artefacts, while allowing relatively low SNR detections to be included. The total number of photometric transients that pass these requirements is roughly 30,000. We emphasise that not all of these transients are SNe, and certainly not all the SNe have adequate light-curve quality and redshift information to be used for cosmological measurements.

During survey operations, the light curve of each DES transient was also fit with the Photometric SuperNova IDentifier software \textsc{psnid} \citep{2011ApJ...738..162S}, a SN photometric classifier tool based on template fitting techniques. This code provided an estimate of the time of peak brightness and a preliminary classification of the SN type.

\subsection{Spectroscopic followup}
\label{sec:spec_followup}
Spectroscopic redshift information on the DES SN candidates is available from a number of sources:
\begin{itemize}
    \item During the course of the DES survey, a wide range of telescopes was used for the spectroscopic follow-up of DES SN candidates \citep[e.g.,][]{DES_spec}. These spectra provide SN classifications and redshifts based on SN spectral features.\footnote{The list of telescopes used for the spectroscopic follow-up of DES SN candidates includes: the 4-metre Anglo-Australian Telescope, the European Southern Observatory Very Large Telescope, Gemini, Gran Telescopio Canarias, Keck, Magellan, MMT, and South African Large Telescope.}
    \item The same telescope programmes also provide spectroscopic redshift measurements from host galaxy spectral features appearing in the SN spectra.
    \item Using the AAOmega spectrograph on the 3.9-m Anglo-Australian Telescope (AAT), spectroscopic redshifts for thousands of galaxies identified as hosts of DES transients were measured as part of the OzDES programme \citep{2015MNRAS.452.3047Y, 2017MNRAS.472..273C, lidman2020ozdes}. The OzDES survey is the primary source of spectroscopic redshifts in the DES photometric SN sample.
    \item Various external redshift catalogues are available in the literature from spectroscopic surveys in the same fields as those monitored by DES-SN.
\end{itemize}
Each source of spectroscopic redshift introduces different selection effects in the DES-SN sample. We describe how these selection effects are modelled in Section~\ref{sec:specz_efficiency}.

\subsection{Host galaxy association}
\label{sec:gal_ass}

For each DES transient, the most likely host galaxy has been identified using the directional light radius (DLR) method \citep{Sullivan_2006, Gupta_2016} applied to galaxies in the SVA1-COADD GOLD image catalogue \citep{Rykoff_2016}. This catalogue uses data in the DES-SN fields collected during the DES \lq Science Verification\rq\ (SV) survey. Within the OzDES survey, a galaxy identified as the host of a DES transient is spectroscopically observed if the following criteria are satisfied:
\begin{enumerate}
    \item The galaxy has the smallest DLR among all catalogue entries 
     and has $\mathrm{DLR}<7$, is brighter than 24.5\,mag in $r$ band, is not flagged as a star \citep[see][for more details]{DES_deepstacks}, and is not in a catalogue of known variable stars and AGN (the so-called \lq VETO\rq\ catalogue);
    \item At least 30 per cent of the detections of the transient passed \autoscan, the transient has at least one detection with a SNR$>$5 in two filters, and at least one filter with two detections with SNR$>$5;
    \item The transient is not detected in multiple seasons (i.e., it is not a long duration transient such as a superluminous SN, a likely AGN, or a variable star);
    \item The day of peak brightness estimated by \textsc{psnid} fitting lies within a DES season.
\end{enumerate}
This set of criteria defines the list of OzDES targets. If a spectroscopic redshift has already been measured by a published redshift survey, or if a spectroscopic redshift has been measured from galaxy features in a live SN spectrum, the galaxy is assigned a lower priority or not targeted at all. In this analysis, we consider OzDES spectroscopic redshifts measured with a confidence level higher than 95 per cent\footnote{A spectroscopic redshift measured with a confidence level higher than 95 per cent corresponds to a quality flag $Q=3$, see \cite{lidman2020ozdes} section~4 for further details on the OzDES redshift flag scheme.} and, if multiple sources of spectroscopic redshift are available for the same host galaxy, we select the OzDES spectroscopic redshift as the more accurate redshift.

After using these host galaxy associations and measurements in the DES-SN3YR analysis, high-quality depth-optimised coadds have been published by \citet{DES_deepstacks}. These coadds have been built combining the highest quality DES-SN images taken before and well after SN detection, with a limiting magnitude of $g\sim27$\,mag, around 1--1.5\,mag deeper then the SV data. As discussed by \citet{DES_deepstacks}, the host galaxy association was revised when upgrading from SV data to the deeper coadds: $\simeq1.1$ per cent of SNe matched to a potential host in SV data had a different host identified with the new coadds. We use these revised associations, and all host galaxy photometric properties are determined from the \citet{DES_deepstacks} stacks. In this paper, we define the host galaxy apparent magnitudes, $m^\mathrm{host}$, as the Kron-like \texttt{MAG\char`_AUTO} magnitudes measured with \textsc{SExtractor} \citep{1996A&AS..117..393B} from the deep coadds.

We identify 7,697 galaxies that satisfy the OzDES selection cuts listed above. 
For 5,049 galaxies, we have a secure redshift measurement.
Table~\ref{table:OzDES_observations} contains a summary of the sources of redshifts.

\begin{table}
\centering
    \caption{Summary of redshift sources for DES SNe.}
    \label{table:OzDES_observations}
    \begin{tabular}{lcc}
 \hline
      Redshift source &   SN redshifts  & \% of Total \\ 
 \hline
All & 5049 & - \\
    \hline
OzDES    &  4419 &  87.52 \\
Galaxy features in SN spectra   &    65 &   1.29 \\
External Catalogues  &   565 &  11.19 \\
\hline
SDSS          &   136 &   2.69 \\
VIPERS        &   105 &   2.08 \\
2dF archival redshifts$^{a}$    &   101 &   2.00 \\
GAMA          &    99 &   1.96 \\
NED           &    32 &   0.63 \\
PanSTARRS+MMT  &    31 &   0.61 \\
ACES  &    19 &   0.38 \\
Others$^{b}$ &     42 & 0.83\\
    \hline
SN features in SN spectra$^{c}$ & 81 & -- \\
 \hline
    \end{tabular}
    \begin{tablenotes}\footnotesize
        \item $^{a}$ Archival redshifts from DEVILS, LADUMA and PanSTARRS SN survey.
        \item $^{b}$ Other external catalogues include  VIMOS VLT Deep Survey (VVDS), ATLAS, MUSE, Ultra Deep Survey (UDS).
        \item $^{c}$ SNe for which the \emph{only} source of spectroscopic redshift is the SN spectrum itself, and either a faint host ($m^\mathrm{host}>24$ for 26 SNe) or no host (55 SNe, \lq hostless\rq SNe) is detected in the deep coadds. These events are excluded from our analysis.
        \item \emph{References:} \citet{VIMOS, DEEP, DEEP2, VIPERS, SAGA, MUSE, 2df, GAMA, ATLAS, ZFIRE, SDSS_DR16, SpARCS, VVDS, UDS, DEVILS, Jones_2018_II, LADUMA}.
    \end{tablenotes}
\end{table}
\subsection{SALT2 fitting and selection cuts}
\label{sec:salt2_data}

To standardise the SNe~Ia brightnesses, the light curves of DES transients with an identified host galaxy and spectroscopic redshift are fit with the SALT2 light-curve model \citep{Guy_2007, Guy_2010}. SALT2 fits provide an estimate of the epoch of SN peak brightness $t_0$, a stretch-like parameter $x_1$, a colour parameter $c$ and the normalisation parameter $x_0$. SALT2 model fitting is implemented with the \snana\ light-curve fitting programme and uses the $\chi^2$ minimization algorithm \textsc{MINUIT} to estimate the best-fitting value and uncertainty of each SALT2 parameter.
The SALT2 parameters are then used to estimate the SN distance modulus, $\mu_\mathrm{obs}$, defined as \citep[e.g.][]{1998A&A...331..815T, 2006A&A...447...31A}:
\begin{equation}
    \mu_\mathrm{obs} = m_B + \alpha x_1 - \beta c + \mathcal{M}_B
    \label{eq:tripp}
\end{equation}
where $m_B$ is defined as $-2.5 \log_{10}(x_0)$ and  $\mathcal{M}_B$ is the absolute brightness for a SN Ia with $x_1=0$ and $c=0$. $\alpha$ and $\beta$ are global nuisance parameters that \lq standardise\rq\ the SN Ia brightnesses, usually determined from a global fit of the Hubble diagram. The residuals from a cosmological model $\Delta\mu$ (often termed \lq Hubble residuals\rq) are
then defined as
\begin{equation}
    \Delta\mu=\mu_\mathrm{obs}-\mu_\mathrm{theory}(\mathcal{C}, z)
    \label{eq:hubble_residual}
\end{equation}
where $\mu_\mathrm{theory}$ is the theoretical distance modulus, which is dependent on the cosmological parameters, $\mathcal{C}$.

In our analysis, we assume $\mathcal{M}_B=-19.365$ and we set $\alpha$ and $\beta$ equal to the values measured by \citet{DES_abbott_ref}, i.e., $\alpha=0.146$, $\beta=3.03$. For both observed and simulated SNe, we measure SN distance moduli, $\mu_\mathrm{obs}$, fixing these nuisance parameters. The values of $\alpha$ and $\beta$ found by \citet{DES_abbott_ref} are also used as the input values for the simulations. We calculate Hubble residuals assuming a flat $\Lambda$CDM cosmological model with Hubble constant $H_0=70$\,km\,s$^{-1}$\,Mpc$^{-1}$ and $\Omega_M=0.311$ \citep[following][]{collaboration2018planck}. While these Hubble residuals are very useful for evaluating our simulations, we note that they do not have the level of accuracy required for a cosmological measurement for several reasons: they are measured from \DiffImg\ photometry, we have not included bias corrections for the SN population, we have not included SN systematic uncertainties, and therefore we have not optimised the values of $\alpha$ and $\beta$. 

To ensure meaningful light-curve fits with the SALT2 model the following selection requirements are applied: i) two filters with at least one epoch with SNR$>$5, ii) at least one data point before the time of peak brightness $t_0$, and iii) at least one data point ten days after $t_0$. Out of 5,049 transients with a host galaxy redshift, 3,627 satisfy these criteria and are successfully fit with the SALT2 model.

This sample of events includes a significant fraction of transients that are clearly not SNe~Ia or core collapse SNe (e.g., AGN, variable stars, or long duration transient events). We use the \lq transient\_status\rq\ flag defined by \citet{DES_spec} to identify multi-season transients, which removes 226 events. Finally, we visually inspect all the remaining transients, and remove artefacts and events that show long term variability (removing an additional 599 events). These single-season requirements reduce the sample to 2,802 visually confirmed SN-like events. 

After light-curve fitting, we consider two sets of additional requirements based on the fitted SALT2 parameters:
\begin{enumerate}
\item \lq Loose\rq\ SALT2-based cuts ($x_1 \in [-4.9,4.9]$ and $c \in [-0.49,0.49]$).
This set of cuts intentionally enhances contamination in the data, and therefore allows us to better analyse the properties of contamination in our sample. After applying these cuts, 249 additional SNe are rejected from the sample (i.e., 2,553 SNe remain);
 \item  The set of SALT2 cuts applied by \citet{Betoule_2014} and \citet{Jones_2017_I} ($x_1 \in [-3,3]$, $c \in [-0.3,0.3]$, $\sigma_{x_1}<1$, $\sigma_{\text{peakMJD}}<2$ days, and fit probability $>$0.001).\footnote{Fit probabilities are based on the fit reduced $\chi^2$ and quantify how well each light curve is described by the SALT2 model assuming the photometric uncertainties are Gaussian.} These cuts are generally adopted in SN Ia cosmology analyses to control contamination from peculiar SNe~Ia or other peculiar thermonuclear SNe that are not well described by a SALT2 model. This set of cuts reduces the data to 1,606 SNe (approximately 30 per cent of the sample is rejected). 
\end{enumerate}
In Table~\ref{table:DES_cuts} we report a summary of the various cuts.

\begin{table}
\centering
    \caption{DES photometric SN sample: summary of data cuts.}
    \label{table:DES_cuts}
\begin{tabular}{lcc}
 \hline
Data cut &   Number & Number \\ 
&   remaining & rejected \\ 
\hline
SNe associated with a spectroscopic redshift & 5049$^{a}$& \\
Fit by SALT2 & 3627$^{b}$ &\\
\lq transient\_status\rq\ flag & 3401 & 226\\
Visual inspection & 2802 & 599$^{c}$\\
Loose SALT2-based cuts & 2553 & 249\\
SALT2-based cuts from \citet{Betoule_2014} & 1606 & 947\\
\hline
    \end{tabular}
    \begin{tablenotes}\footnotesize
        \item $^{a}$ Including 54 SNe/hosts located in the DECam inter-CCD chip gaps;
        \item$^{b}$ We exclude events for which the redshift is estimated from SN spectral features in the SN spectrum; 
        \item $^{c}$ Out of the 599 visually inspected events, only 112 would pass the loose SALT2 cuts and only eight would pass the \citet[][]{Betoule_2014} SALT2-based cuts.  
    \end{tablenotes}
\end{table}

\begin{figure*}
    \includegraphics[width=0.99\linewidth]{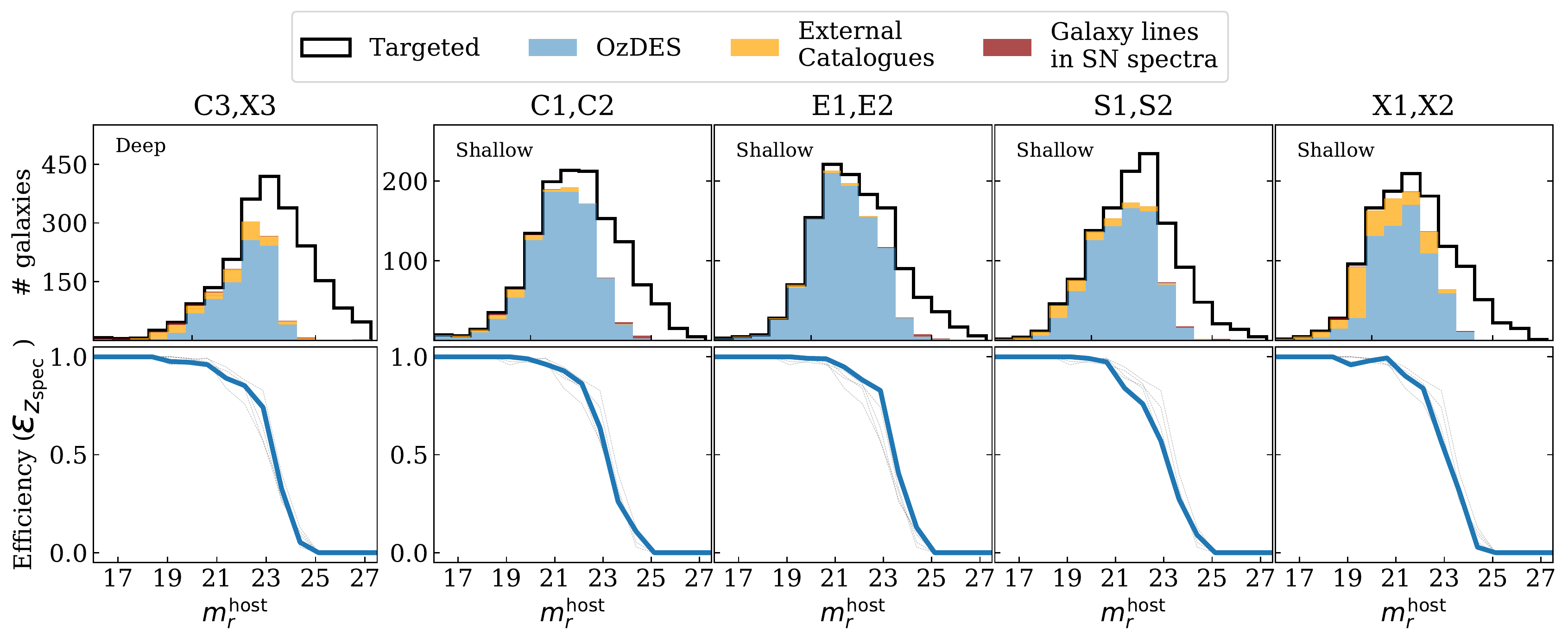}
\caption{\emph{Top panels}: For each pair of DES SN fields we present distributions of $m^\mathrm{host}_r$ for all host galaxies that passed the OzDES selection criteria listed in Section~\ref{sec:gal_ass} (black histogram), for host galaxies with a spectroscopic redshift from OzDES (light blue), from external catalogues (orange), and from galaxy emission lines in SN spectra (dark red). \emph{Bottom panels}: we show \effz\ vs $m_r^\mathrm{host}$ for each pair of SN fields (blue thick line) and for all other SN fields (thin grey lines). The average host galaxy brightness is $m_r^\mathrm{host}\simeq23$\,mag in the deep fields (X3 and C3) and $m_r^\mathrm{host}\simeq22$\,mag in the shallow fields.}
    \label{fig:efficiencies}
\end{figure*}

\section{Spectroscopic redshift efficiency}
\label{sec:specz_efficiency}

As part of a SN~Ia cosmology analysis, modelling selection effects is essential to estimate bias corrections and simulate training samples. Detection efficiency and photometric instrumental effects for the DES SN program have been characterized and presented by \citet{2015AJ....150..172K}. In this analysis, we mainly focus on selection effects due to the requirement of a host galaxy spectroscopic redshift. This is a critical selection effect in the DES SN dataset -- it shapes the redshift distribution of the sample and introduces biases towards SNe in bright, emission line galaxies for which measuring a spectroscopic redshift is easier. 

In this section, we describe our approach for the modelling of the spectroscopic redshift efficiency (\effz), i.e., the overall efficiency of obtaining spectroscopic redshifts in DES and how we incorporate this in our simulations of the DES SN sample. 

\subsection{A novel approach to modelling selection effects}

Previous analyses of photometric SN samples \citep{Jones_2017_I, Jones_Foundation} have modelled \effz\ as a one-dimensional function of redshift, tuning \effz\  so that the simulations reproduce the observed redshift distribution. By construction, this efficiency function is tailored to a specific choice of volumetric SN rates, it does not depend on galaxy properties, and it is applied to all types of SNe. While this approach guarantees a good agreement in the redshift distribution between data and simulations, it does not account for brighter galaxies being more likely to get a spectroscopic redshift and, as a consequence, that SNe exploding in bright and high mass galaxies are more likely to be selected. 

Our approach is substantially different in two respects. First, we measure \effz\ from the data -- the sample of host galaxies that satisfy the criteria listed in Section~\ref{sec:gal_ass}, and therefore have been targeted in the OzDES survey. Second, we measure \effz\ as a function of SN host galaxy properties. Using the sample of targeted galaxies, we calculate the fraction of galaxies with and without a spectroscopic redshift and measure the efficiency as a function of the host galaxy brightness and other observables, including the host galaxy $g-r$ colour and the epoch of SN discovery.

Our efficiency function can be integrated into simulations, but it in turn requires the simulations to include host galaxies with realistic properties. 
In particular, our simulations need to account for the strong dependence of SN rates on galaxy properties (for a given SN, not every galaxy is equally likely to be the host galaxy, depending on the galaxy stellar mass and/or the galaxy star formation rate). Using empirical SN rate models, the simulated host galaxies should reproduce the properties and brightness distributions of the observed SN host galaxies. This approach is fundamentally data driven, and takes into account the fact that different types of SNe explode in different populations of galaxies with different brightness distributions.

In this implementation, a good match between simulations and data is not guaranteed, as none of the parameters is tuned to ensure this. Our method also enables a novel independent astrophysical test of whether measurements of SN rates and their dependencies on galaxy properties are well understood across the redshift range covered by the DES SN sample. 

\subsection{Efficiency of the spectroscopic redshift survey}
\label{sec:efficiency}

Spectroscopic redshifts are available from various sources (Section \ref{sec:spec_followup}), primarily from host galaxy spectral features and, when the live SN spectrum is available, from SN spectral features. When the redshift is measured from galaxy spectral features, \effz\ depends primarily on the brightness of the host galaxy and the host spectral type. For a subset of 81 of the spectroscopically confirmed SNe (Table~\ref{table:OzDES_observations}), the redshift can only be estimated from SN spectral features, and \effz\ depends on the brightness of the SN on the epoch of spectroscopic observation. Therefore, including SN events for which the \emph{only} source of redshift is from the SN spectral features would require a very different and independent selection function \citep[e.g. the selection functions presented in][]{DES_biascor, DES_spec}. This is beyond the scope of this analysis, and we therefore exclude this redshift information from this paper.

We measure \effz\ as a function of host galaxy brightness (Section \ref{sec:eff_brightness}), host galaxy observed colour (Section \ref{sec:eff_colour}) and the year of discovery of the SN (Section \ref{sec:eff_time}). We define the efficiency as the ratio of the number of host galaxies for which a redshift is available (either from OzDES or other catalogues), over the total number of host galaxies that passed OzDES selection criteria. The OzDES selection criteria are listed in Section~\ref{sec:gal_ass}, which are different from the selection cuts used to define the final DES photometric SN sample (Section~\ref{sec:salt2_data}).

\subsubsection{Efficiency as a function of galaxy brightness}
\label{sec:eff_brightness}

We first measure \effz\ as a function of $m^\mathrm{host}_r$, presented in Fig.~\ref{fig:efficiencies} for five sub-groups of DES SN fields. As expected, \effz\ is high for bright host galaxy magnitudes, in many cases 100 per cent, and drops sharply above $m^\mathrm{host}_r\sim21$\,mag. The 50 per cent efficiencies range from $m^\mathrm{host}_r\simeq$ $23$ to $23.5$\,mag.

The efficiency varies from field to field for several reasons. Firstly, the two deep fields, X3 and C3, were prioritised by OzDES as they include more SN candidates due to the deeper DES data. Secondly, the E1 and E2 fields were observed more frequently as they have the longest visibility window from the AAT. Finally, some fields have more external redshifts available; for example, the X1 and X2 fields overlap with the GAMA survey \citep{GAMA}.

\begin{figure*}
\centering
    \includegraphics[width=0.95\linewidth]{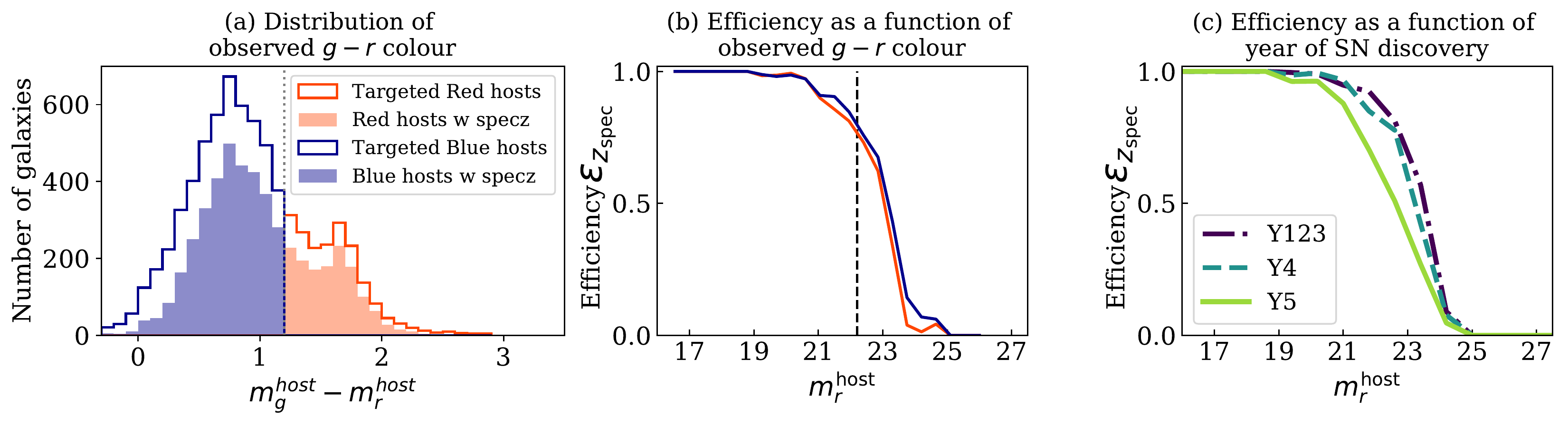}
    \caption{Panel \textbf{(a)}: distribution of observed $g-r$ colours for all host galaxies that passed the OzDES targeting criteria (open histogram) and hosts with spectroscopic redshift (filled histograms). We define red galaxies as those with $g-r$ greater than 1.2, and blue galaxies as those with $g-r$ less than 1.2 (marked by the vertical dotted line). Panel \textbf{(b)}: \effz\ versus $m^{\mathrm{host}}_r$ for both red and blue galaxies (red and blue solid lines). The median brightness of our sample of SN hosts is shown as the vertical dashed line, and it shows the magnitude at which most of the DES host galaxies are observed and therefore where discrepancies in efficiency have the largest impact. Panel \textbf{(c)}: \effz\ as a function of $m^{\mathrm{host}}_r$ for SNe discovered in the first, second and third year of DES (2013--2015; dotted-dashed line), for SNe discovered in the fourth year of DES (2016; dashed line) and in the fifth year of DES (2017; solid line). The limited observing time towards the end of the OzDES program caused a small drop in \effz\ for Y4 and Y5.}
    \label{fig:eff_colour}
\end{figure*}

\subsubsection{Efficiency as a function of galaxy spectral type}
\label{sec:eff_colour}

\effz\ depends not only on galaxy brightness but also on the galaxy spectral type (e.g., it is easier to measure redshifts for emission line galaxies). This dependence affects the fraction of core collapse SN contamination in our sample as these events almost exclusively explode in star forming galaxies \citep{2011MNRAS.412.1441L}.
Since the spectral type is not available for all the targeted host galaxies, we consider alternative proxies of galaxy spectral type, such as the observed $g-r$ colour.

In Fig.~\ref{fig:eff_colour}, we present the distribution of observed $g-r$ colours for the sample of SN host galaxies that pass the OzDES criteria (see Section~\ref{sec:gal_ass}). We separately measure \effz\ for the 25 per cent \lq reddest\rq\ galaxies in the sample and for the remaining sample of \lq bluer\rq\ galaxies (this corresponds to a threshold of $g-r = 1.2$\,mag). The efficiency measured from the sub-sample of \lq redder\rq\ galaxies is systematically lower than that measured from \lq bluer\rq\ galaxies (5 per cent lower at $m_r^{host}=$22\,mag and 15 per cent lower at $m_r^{host}=$23\,mag). We implement this colour-dependency of \effz\ in our simulations. We note that this colour-dependency is a second-order effect as the OzDES programme is optimised to achieve a high completeness to a magnitude limit of $m_r^{\mathrm{host}}24$ and the OzDES strategy is to repeatedly target SN host galaxies until the level of confidence is larger than 99 per cent \citep[see][for details]{lidman2020ozdes}. 

\subsubsection{Efficiency as a function of the year of SN discovery}
\label{sec:eff_time}

The OzDES programme ran between 2013 (first year of the DES SN programme) and 2018 (one year after the end of the DES SN programme), so that host galaxies of SN discovered in the last year of DES could be observed. The number of nights allocated to OzDES was progressively increased each year \citep[see][for details]{lidman2020ozdes} in order to accommodate the increasing number of SNe discovered by DES. The amount of fiber hours available at the end of OzDES was not sufficient to achieve the same efficiency obtained for hosts of SNe discovered earlier in the DES survey.
For this reason, we find that \effz\ decreases for SNe discovered in the fourth and fifth years of DES. This trend is shown in Fig.~\ref{fig:eff_colour}\emph{(c)} and is modelled in our simulations for shallow and deep fields separately.

\section{Simulations}\label{sec:simulations}

We next describe the simulations that underpin our study of the systematic uncertainties introduced by contamination from core collapse SNe. These simulations are designed to produce a realistic realisation of the DES photometric SN sample. In the following section we present the \lq Baseline\rq\ simulation based on assumptions about the global properties of SNe Ia, peculiar SNe Ia and core collapse SNe. In Section~\ref{sec:sim_cc_vary}, we present additional simulations and explore alternative core collapse SN modelling assumptions.

\subsection{Implementation in SNANA}
\label{sec:snana_sim}
Synthetic SN light curves are generated and analysed using the SuperNova ANAlysis software \citep[\snana,][]{Kessler_2009},\footnote{\url{https://github.com/RickKessler/SNANA}} integrated in the \textsc{pippin} pipeline framework \citep{Hinton2020}.\footnote{\url{https://github.com/Samreay/Pippin}}
The \snana\ simulation generates realistic transient light curves from one or more spectrophotometric models of transients. \citet[][hereafter \citetalias{DES_biascor}]{DES_biascor} present a detailed description of the simulations designed to characterise and reproduce SNe Ia within the DES SN survey, and in particular the DES-SN3YR sample. Here we briefly describe the three main steps that constitute the \snana\ simulation (see figure 1 in \citetalias{DES_biascor} for a schematic illustration) and highlight the assumptions adopted in our analysis.

The first step is to generate a source SED model, selecting a specific SN population (see Section~\ref{sec:sim_ia}, \ref{sec:sim_pecia} and \ref{sec:sim_cc_baseline}) and astrophysical effects that include host galaxy extinction, redshifting, cosmological dimming, lensing magnification, peculiar velocity and Milky Way extinction. In our analysis, we use where necessary a \cite{1989ApJ...345..245C} dust law with $R_V=3.1$ for Milky Way and host galaxy dust extinction. The integration of the generated SED model over the DES filters provides an estimate of the \lq true\rq\ magnitudes of the source before observational noise is applied.

The second step is to convert true magnitudes into observed fluxes and calculate the flux uncertainties. This step uses the observing conditions provided in a pre-computed observational library (referred to as a \lq \texttt{simlib}\rq). The \texttt{simlib} includes measured photometric zero-points, sky noise and point spread function (PSF) information at 10,000 random sky locations within the DES fields. Flux uncertainties are estimated as the quadrature sum of the sky noise and the Poisson noise from the source and the surface brightness of the host galaxy. Host galaxies are selected from a galaxy catalogue (\lq \texttt{HOSTLIB}\rq). In Section~\ref{sec:sim_hostgalaxies}, we present the \texttt{HOSTLIB} used for our simulations and the recipe implemented for host galaxy association. Finally, the extra source of anomalous noise introduced by the \DiffImg\ pipeline is estimated and robustly modelled using a set of separate image-based simulations for which \lq fake\rq\ SNe are placed in real DES images and processed through the same \DiffImg\ pipeline as applied to the data (see \citet{2015AJ....150..172K} and section 6.4 in \citetalias{DES_biascor} for an extended discussion).

The third and final step is to simulate the \lq trigger model\rq\ for the selection of events. Detection efficiency versus signal-to-noise ratio is implemented as described in section 7.1 in \citetalias{DES_biascor}. Following the same DES trigger logic applied to real data, we select simulated events that have at least one detection on two separate nights.

In the following subsections we describe the SED models used to simulate different astrophysics transients and their implementation in the simulation. 

\subsection{Simulations of \lq normal\rq\ SNe~Ia}\label{sec:sim_ia}

We simulate normal SNe~Ia, i.e., those that are used in cosmological fitting, using the SALT2 SED model presented by \citet{Guy_2007} and trained on the Joint Lightcurve Analysis sample presented by \citet{Betoule_2014}.
Each SN Ia is generated with random redshift, $t_0$, $x_1$ and $c$ values. Redshifts are generated following the volumetric rate presented by \citet{Frohmaier_2019}, who combined published measurements from \citet{Dilday_2008} and \citet{Perrett_2012} with new measurements from the Palomar Transient Factory \citep[PTF;][]{PTF_REF}. The $t_0$ are randomly distributed within a time window that starts two months before the beginning of DES and finishes two months after the last visit of DES to the SN fields. The underlying distributions of $x_1$ and $c$ are taken from \citet{Scolnic_2016}.
For SN Ia intrinsic scatter, we adopt the \lq G10\rq\ spectral variation model from \cite{Kessler_2013} that is based on the wavelength-dependent scatter presented by \cite{2010A&A...523A...7G}. Future analyses will explore in greater depth other approaches to simulating SNe~Ia in DES, including different intrinsic scatter models \citep{BS20} and various effects of correlations between SNe~Ia and host galaxy properties \citep{Sullivan_2006, Smith_2012,rigault2018strong, DES_massstep}. In this analysis, the only SN Ia-host correlation that we model is between $x_1$ and host galaxy stellar mass (see Section~\ref{sec:sim_hostgalaxies} for details).

\subsection{Simulations of peculiar SNe~Ia}
\label{sec:sim_pecia}
\begin{figure*}
  \centering
  {\includegraphics[width=0.9\linewidth]{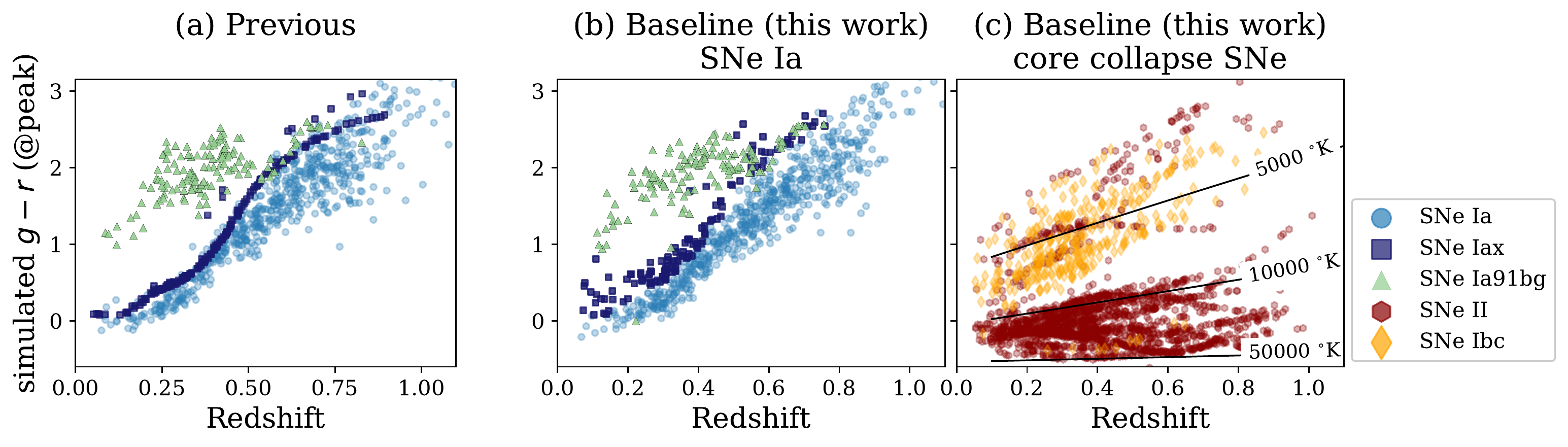}}
\caption{Simulated $g-r$ colours at peak brightness for various types of SNe as a function of redshift. In panel \textbf{(a)} SNe~Ia are generated using the SALT2 SED model from \citet{Betoule_2014}, and SNe 91bg and SNe~Iax using the original PLAsTICC templates. In panel \textbf{(b)} SNe~Ia are generated from SALT2 SED model, and SNe 91bg and SNe~Iax are simulated using the PLAsTICC templates with the addition of dust extinction for SNe~Iax and stretch diversity for SNe 91bg (see Section~\ref{sec:sim_pecia}). In panel \textbf{(c)}, core collapse SNe are simulated using the \citetalias{Vincenzi_2019} templates that include dust extinction as measured in the original events. This is the baseline simulation implemented in this analysis. For comparison with the $g-r$ colour evolution of core collapse SNe, we also show the $g-r$ colour measured from black body SEDs at temperatures of 5000, 10000 and 50000\,K.}
  \label{fig:gr_colour_peak_iaPecia}
\end{figure*}

\begin{figure}
  \centering
  \subfigure[]{\includegraphics[width=0.8\linewidth]{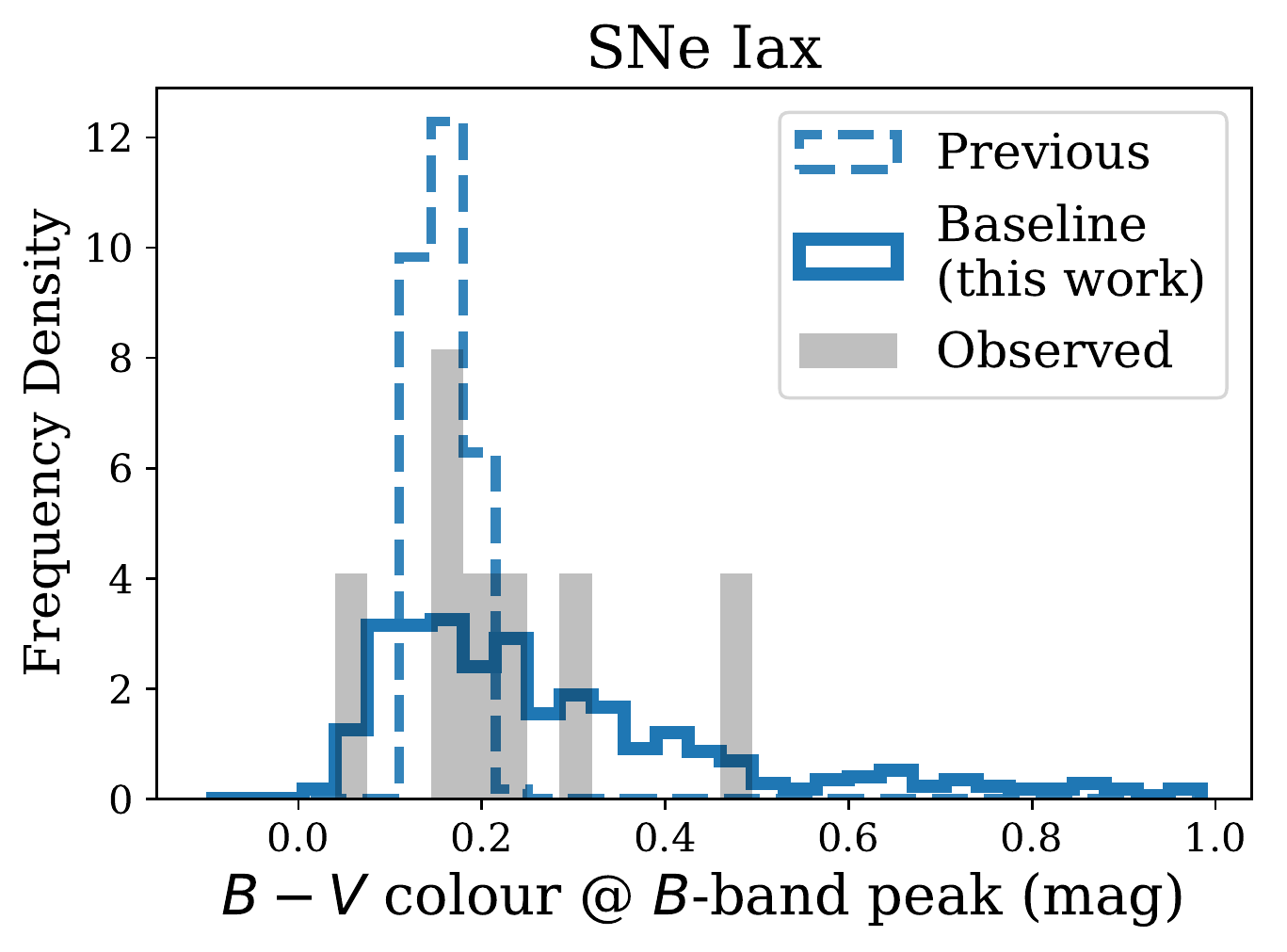}}\quad
\caption{$B-V$ colour distribution at $B$-band peak for SNe~Iax simulated using the original PLAsTiCC models (dashed histogram), for the updated SNe~Iax model used in this analysis (solid histogram; Section~\ref{sec:sim_pecia}) and for seven observed SNe~Iax with for which good $B$ and $V$ band photometry around peak has been published \citep[grey-filled histogram, SN~2003gq, SN~2005cc, SN~2005hk, SN~2008A, SN~2008ha,  SN~2011ay, SN~2012Z from ][]{2012MNRAS.425.1789S, Foley_2013, 2019MNRAS.490.3882S}.}
  \label{fig:sneiax}
\end{figure}

We include in our simulations two types of peculiar SNe~Ia that may appear as photometric contaminants in SN~Ia samples: SN1991bg-like SNe \citep{1992AJ....104.1543F} and SN2002cx-like supernovae \citep[][hereafter SNe~Iax]{Li_2003, Foley_2013}. SN1991bg-like (\lq 91bg-like\rq) SNe are sub-luminous compared to normal SNe~Ia, and characterised by fast-declining (small $x_1$), light curves and redder colours at peak.
In our simulations, we use the SED library of 35 91bg-like events presented in PLAsTiCC \citep{2019PASP..131i4501K}. In the original PLAsTiCC simulation, only five different SEDs were used and no stretch diversity was simulated \citep[see Section 4.2.2 in][]{2019PASP..131i4501K} due to an error in the generation of the models. For our simulations the PLAsTiCC team have provided us with the correct set of SED models. In Fig.~\ref{fig:gr_colour_peak_iaPecia} we present the $g-r$ colour synthesised at peak before observational noise is applied for our simulated 91bg-like SNe. This sub-class of peculiar SNe~Ia is significantly redder at peak compared to normal SNe~Ia. 

SNe~Iax \citep[see][for a recent overview]{2017hsn..book..375J} generally rise and decline faster than normal SNe~Ia and are characterised by low-velocity ejecta. Again, we use the model presented in PLAsTiCC, based on SN\,2005hk \citep{Phillips_2007, Sahu_2008}. As with normal SNe~Ia, the absolute brightness of SNe~Iax has been shown to be correlated with light-curve width \citep{Foley_2013}. To reproduce this correlation and expand the diversity of SN Iax models, the PLAsTiCC team generated multiple SN Iax SEDs by warping and renormalising the original SN\,2005hk template.
This reproduces the diversity of SNe~Iax in terms of light curve shape and normalisation, but leaves the colour properties at peak unchanged (see Fig.~\ref{fig:gr_colour_peak_iaPecia} and \ref{fig:sneiax}). The colour evolution and scatter of SNe~Iax are poorly understood. However, as SNe~Iax are believed to explode in younger environments \citep{Takaro_2020}, and are therefore likely to be affected by dust, we opt to use dust extinction to introduce variation in the colour of the models. The reddening within the host galaxy for SN\,2005hk is estimated to be $E(B-V)=0.09$ \citep{Chornock_2006}, so we correct the PLAsTiCC SN Iax models for $E(B-V)=0.09$, and apply a range of host extinctions in the simulations. We adopt the host extinction distribution described by \citet{Rodney_2014} (which we also adopt for core collapse SNe in the following sections), which allow us to well reproduce the colour diversity observed for SNe~Iax (see Fig.~\ref{fig:sneiax}).

Our revision of the original PLAsTiCC SN Iax models addresses the issues identified by \citet{Popovic_2020}. They included the PLAsTiCC SN~Iax models in their simulations of the Sloan Digital Sky Survey (SDSS) photometric SN sample, and observed that this significantly overestimates the predicted contamination, with the simulated SNe~Iax appearing bluer than other samples of observed SNe~Iax (see Fig.~\ref{fig:sneiax}).

\subsection{Simulations of core collapse SNe: baseline approach}
\label{sec:sim_cc_baseline}

Our Baseline core collapse SN simulations use the library of 67 SED time-series templates presented by \citet[][hereafter \citetalias{Vincenzi_2019}]{Vincenzi_2019}. This library combines spectroscopy and multi-band photometry from 67 well-observed core collapse SNe across 6 different subclasses (SN II, SN IIb, SN IIn, SN Ib, SN Ic and SN Ic-BL). Each template covers 1600--11000\AA; the UV coverage, in particular, is critical when simulating core collapse SNe at high redshift. Fig.~\ref{fig:gr_colour_peak_iaPecia} shows the redshift evolution of the simulated $g-r$ colour at peak for different types of core collapse SNe compared to SNe~Ia. We find that core collapse events in our simulations have the expected colour evolution. Stripped-envelope SNe are systematically redder at peak compared to SNe~Ia. SNe~II, however, are significantly bluer events and they follow the colour evolution expected from black body SEDs at different temperatures. 

By construction the \citetalias{Vincenzi_2019} template library is biased towards bright core collapse SNe and may not be representative of the intrinsic brightnesses and relative rates of different sub-types. Luminosity distributions and relative rates are generally measured from magnitude-limited samples such as the Lick Observatory Supernova Survey sample \citep[LOSS,][]{Leaman_2011, 2011MNRAS.412.1441L}. As the SN events in the LOSS sample do not have sufficient data quality to construct SED templates, we adopt a hybrid approach and use the biased sample of SN events in the \citetalias{Vincenzi_2019} template library and normalise it to brightnesses and rates measured from the LOSS sample.

For core collapse SN relative rates, we use the measurements presented by \citet{Shivvers_2017}. Using the LOSS sample and revising the \citet{2011MNRAS.412.1441L} measurement, \citet{Shivvers_2017} showed that in the local universe SNe~II and stripped-envelope SNe represent 69.6 per cent and 30.4 per cent of all core collapse SNe respectively. \citet{2020arXiv201015270F} find a similar result using data from PTF. Given the lack of measurements of relative rates at higher redshifts, in our Baseline simulation we assume that these relative rates do not evolve with redshift.
We simulate core collapse SNe assuming the rate follows the cosmic star formation history presented in \citet{Madau_2014} normalised by the local SN rate of Frohmaier et al. 2020. 

For the luminosity functions, the baseline simulation uses the mean and r.m.s absolute brightnesses measured from the LOSS sample, and we interpret these measurements as Gaussian luminosity functions. These were revised in \citetalias{Vincenzi_2019} following updated classifications published by \citet{Shivvers_2017} and they are reported in Table~\ref{table:LFs}. We use the set of \citetalias{Vincenzi_2019} templates that has not been corrected for host-galaxy dust extinction because the revised \citet[][hereafter \citetalias{2011MNRAS.412.1441L}]{2011MNRAS.412.1441L} luminosity functions are also measured from SNe not corrected for host-galaxy dust extinction. As described by \citetalias{Vincenzi_2019}, each sub-type of template is matched to its respective luminosity function applying \emph{sub-type dependent} magnitude offsets and dispersion.

The simulated core collapse SN contamination can vary significantly depending on the choice of luminosity function, on whether additional host extinction is simulated, and on the adopted distribution of host-galaxy dust extinction. As most of these quantities are poorly constrained (especially at high redshift) we do not rely on one single core collapse SN simulation but instead design a set of simulations that explore these different assumptions, and we test how our modelling choices affect our analysis. In Section~\ref{sec:sim_cc_vary}, we present in detail each core collapse simulation built for this analysis.

\subsection{Simulating host galaxies}
\label{sec:sim_hostgalaxies}
\begin{figure*}
  \centering
\includegraphics[width=0.99\textwidth]{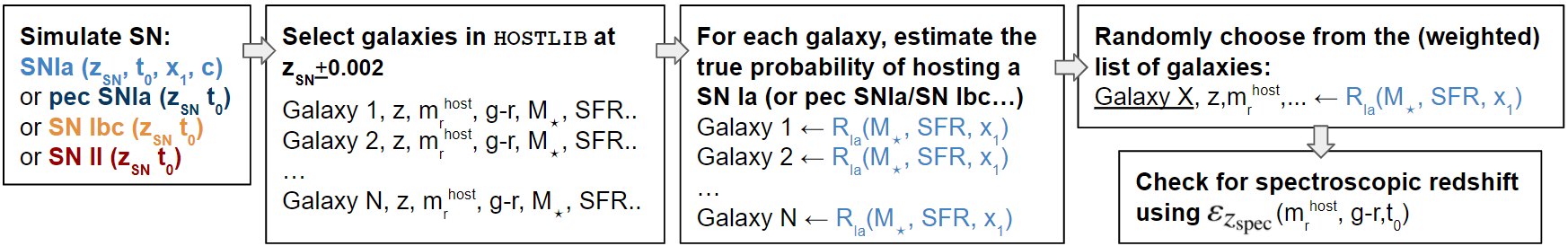}
  \caption{Flow chart describing the host galaxy association in the \textsc{SNANA} simulations. Here we show an example of host galaxy association for SNe~Ia, but the same general process applies to other SN types. Equations~\ref{eq:rate_ia}, \ref{eq:rate_ia2} and \ref{eq:rate_cc} in Section~\ref{sec:host_rate_ia} and \ref{sec:host_rate_cc} describe SN rates as a function of galaxy properties (and additionally $x_1$ for SNe~Ia) for all the SN types included in our simulations.}
  \label{fig:host_scheme}
\end{figure*}
\begin{figure}
  \centering
\includegraphics[width=0.49\textwidth]{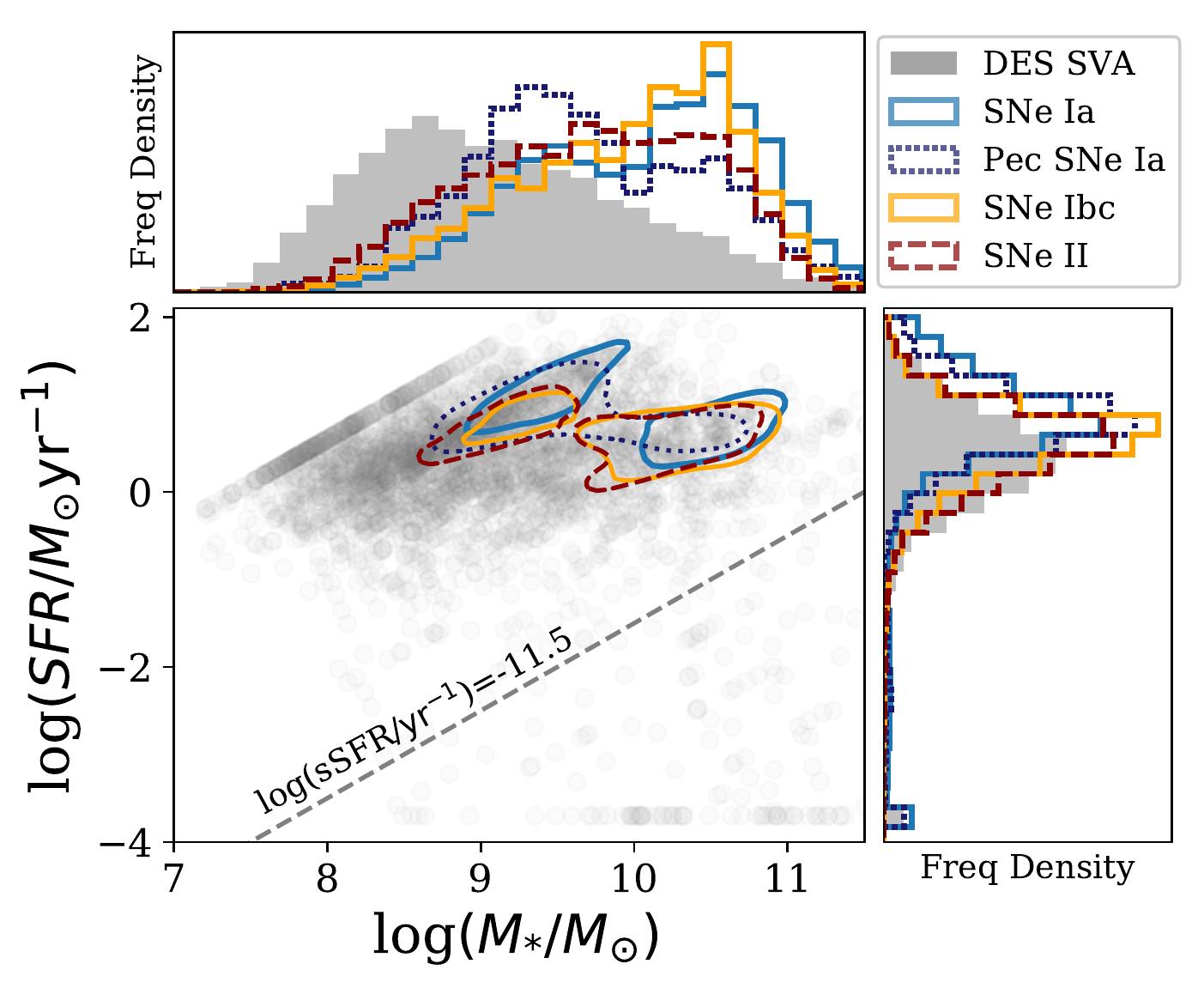}
  \caption{Distribution of galaxy SFR ($\log(\mathrm{SFR})$) versus galaxy stellar mass ($\log(M_*)$) for all galaxies the \texttt{HOSTLIB} (grey symbols anf filled grey histograms) and for four different types of simulated SNe: SNe~Ia ( solid blue line), peculiar SNe~Ia (91bg-like and SNe~Iax; dotted blue line), SNe~Ibc (solid orange line) and SNe~II (dashed red line). The central 2D plot shows the 68 per cent density contour for each SN type. Different types of SNe are associated to host galaxies following the SN rates presented in Section~\ref{sec:sim_hostgalaxies}. The dashed grey line separates our definition of star forming (above the line, i.e., $\log(\textrm{sSFR})>-11.5$) and passive galaxies (below the line, i.e., $\log(\textrm{sSFR})<-11.5$).}
  \label{fig:host_sims}
\end{figure}

The rates of SNe in galaxies depend on the galaxy properties, such as  stellar mass ($M_{*}$), star formation rate (SFR), and metallicity \citep{Sullivan_2006, Lampeitl_2010, 2011MNRAS.412.1441L, Smith_2012, 2013MNRAS.435.1680J, Graur_2015, rigault2018strong, Graur_2017}. For any given SN type, not every galaxy is equally likely to be a host and, in addition, the likelihood of a SN host having a spectroscopic redshift depends on the galaxy properties (see Section~\ref{sec:efficiency}). Therefore, realistic simulations require an accurate modelling of how the SN rate and \effz\ are correlated with galaxy properties. In this section we discuss our approach in the simulations. A schematic illustration of galaxy association is presented in Fig.~\ref{fig:host_scheme}.

\subsubsection{Simulating host galaxies of SNe~Ia}
\label{sec:host_rate_ia}

We model correlations between SN Ia rates and galaxy properties following a two-component parametrization (the \lq A+B\rq\ model) introduced by \citet{Mannucci_2005}. In this approach, the SN Ia rate is described as the sum of two terms:
\begin{equation}
R^{A+B}_{\mathrm{Ia}}\left(M_{*},\mathrm{SFR}\right) = A \times M_{*} + B \times \mathrm{SFR} \\
\label{eq:rate_ia}
\end{equation}
This model was implemented by \citet{Sullivan_2006} to analyse the Supernova Legacy Survey (SNLS) SN Ia sample. We use the best-fitting $A$ and $B$ parameters presented by \citet{Sullivan_2006}.

To model the well-known correlation between SN~Ia $x_1$ and host galaxy $M_{*}$ \citep[e.g., figure 4 in][]{DES_massstep}, we multiply the SNLS SN Ia rate in equation~\ref{eq:rate_ia} by an additional term ($R^{*}_{\mathrm{Ia}}(x_1, M_{*})$)\, so that the rate of SNe~Ia in galaxies with $M_{*}<10^{10}$\,$M_{\sun}$ drops monotonically to zero with decreasing $x_1$. After analysing the DES-SN3YR SN~Ia sample and comparing the tail of SNe Ia with $x_1<0$ in high mass galaxies ($M_{*}>10^{10}$\,$M_{\sun}$) and low mass galaxies ($M_{*}<10^{10}$\,$M_{\sun}$), we model the relative probability of having a SNe Ia with a SALT2 stretch $x_1$ in a galaxy with stellar mass $M_{*}$ as: 
\begin{equation}
\begin{split}
R^{*}_{\mathrm{Ia}}(x_1, M_{*}) &= e^{-x_1^2}&\text{ for }x_1<0\text{ and }M_{*}<10^{10}\,M_{\sun}\\
R^{*}_{\mathrm{Ia}}(x_1, M_{*})& = 1 &\text{ for }x_1>0\text{ and }M_{*}<10^{10}\,M_{\sun}\\
R^{*}_{\mathrm{Ia}}(x_1, M_{*})& = 1 &\text{ for }\forall x_1 \text{ and }M_{*}>10^{10}\,M_{\sun}.
\label{eq:rate_ia2}
\end{split}
\end{equation}
As a result, the net rate applied for SNe Ia is:
\begin{equation}
R_{\mathrm{Ia}}(M_{*},\mathrm{SFR}, x_1) \propto R^{A+B}_{\mathrm{Ia}}(M_{*},\mathrm{SFR}) \times R^{*}_{\mathrm{Ia}}(x_1, M_{*}).
\label{eq:rate_ia3}
\end{equation}
For peculiar SNe~Ia we apply the same SN rate model used for normal SNe~Ia with some variations. 91bg-like SNe~Ia primarily explode in E/S0 galaxies \citep{2001ApJ...554L.193H, 2011MNRAS.412.1441L}, while SNe~Iax are rarely found in early-type galaxies \citep{Takaro_2020}. Therefore, we set the rate of 91bg-like (SNe~Iax) to be zero in star forming (passive) galaxies. In our analysis, a galaxy is defined as passive if its specific star formation rate sSFR (the star-formation rate per unit stellar mass) is smaller than $10^{-11.5}$\,yr$^{-1}$ (Fig.~\ref{fig:host_sims}).

\subsubsection{Simulating host galaxies of core collapse SNe}
\label{sec:host_rate_cc}

Core collapse SNe occur almost exclusively in star-forming galaxies \citep{2011MNRAS.412.1441L, Kelly_2012, Graur_2017}. 
\citet{Graur_2017} measured the core collapse SN rate as a function of galaxy properties for stripped envelope SNe and SNe~II respectively. These rates are calculated using core collapse SNe in the LOSS sample and are presented as a function of $M_{*}$, which is correlated with SFR for star forming galaxies. Following these measurements we model core collapse SN rates as:
\begin{equation}
   \begin{split}
    \text{R}_{\text{Ibc/II}} &= 0 \text{ in passive galaxies}\\
    \text{R}_{\text{Ibc}}(M_{*}) & \propto 
    (M_{*}/\text{M}_{\odot})^{0.36}\\
    \text{R}_{\text{II}}(M_{*}) & \propto 
    (M_{*}/\text{M}_{\odot})^{0.16}\\
    \end{split}
    \label{eq:rate_cc}
\end{equation}
\citet{Graur_2017} show that SNe~II have a shallower dependency on $M_{*}$ compared to stripped-envelope SNe, and this result has a statistical significance of $>2\sigma$. This difference implies that the ratio between stripped-envelope SNe and SNe~II \citep[that on average is roughly 0.435; see][]{Shivvers_2017} varies depending on the host galaxy $M_{*}$; stripped-envelope SNe are ten times less common than SNe~II in low-mass galaxies, but almost 1/3 of the SN~II rate in high-mass galaxies. 
At higher redshifts, the DES photometric SN sample is  biased towards brighter and more massive galaxies as they are more likely to get a spectroscopic redshift. This bias affects the composition of core collapse SN contamination as a function of redshift and is modelled in our simulations.

\subsubsection{Host galaxy association in simulations}
\label{sec:host_association}

Following \citet{DES_massstep}, we select SN host galaxies from a \texttt{HOSTLIB} (Section~\ref{sec:snana_sim}) generated from DES SV data. This catalogue includes $\sim$380,000 galaxies for which quantities like redshift (spectroscopic or photometric), galactic coordinates, magnitudes and S{\'e}rsic profiles \citep{1963BAAA....6...41S} have been measured. For each \texttt{HOSTLIB} galaxy, $M_{*}$ and SFR are measured using the method presented by \citet{DES_massstep} (see section 2.2.2).

The completeness of the DES SV \texttt{HOSTLIB} is $>99$ per cent for $m^{\mathrm{host}}_i<23.8$\,mag and 50 per cent for $m^{\mathrm{host}}_i<24.75$\,mag. Analysing the SNLS spectroscopic SN Ia sample \citep{Sullivan_2010}, the fraction of SNe Ia in galaxies fainter than 23.8 is less than 15 per cent for $z<0.8$ and approximately 30 per cent at $z=1$. This fraction is likely to be higher for core collapse SNe that on average explode in fainter galaxies. The depth of the DES SV \texttt{HOSTLIB} is one of the  limiting factors in our analysis and may result in an overestimate of SNe at higher redshifts. We will explore the implementation of deeper \texttt{HOSTLIB} catalogues in future articles.

In our simulations, the SN-to-galaxy association is implemented as follows (see Fig.~\ref{fig:host_scheme} for a schematic illustration). For a SN event simulated at redshift $z$ we select all \texttt{HOSTLIB} galaxies within the interval $z\pm0.002$. Each galaxy within this redshift interval is then weighted by the SN rate (Sections~\ref{sec:host_rate_ia} and \ref{sec:host_rate_cc}), so that high mass galaxies are favoured and the large fraction of faint, low-mass galaxies are given lower weight.  The host is then randomly selected from the weighted list of galaxies. We identify the location of the SN within a host assuming that the distribution of SNe within their host galaxies follows the galaxy light profile \citep{2008ApJ...687.1201K}. For each epoch, the simulation computes the host galaxy flux within the $2\sigma_{\text{PSF}}$ radius aperture from the location of the SN and this Poisson variance is added to the flux variance. This galaxy variance affects the signal-to-noise of the SN flux and its likelihood of being detected.
Finally, given the $m^\mathrm{host}_r$ and $g-r$ colour of the selected host galaxy, as well as the year of discovery of the simulated SN, we apply the efficiency \effz\ (Section~\ref{sec:specz_efficiency}) to determine whether a redshift is measured.

Our method for the simulation of SN host galaxies is a significant improvement over earlier work. Our approach accounts for the fact that SNe of different astrophysical origin occur in different types of galaxies with different rates. Using our baseline simulation, we show in Fig.~\ref{fig:host_sims} how simulated host galaxies of different types of SNe have different distributions in terms of simulated $M_{*}$ and SFR.
Compared to published samples of SNe~Ia and core collapse SNe \citep{DES_deepstacks, ztf_BTS, 2011MNRAS.412.1441L}, our simulations reproduce the observed host galaxy properties: the population of SN~Ia hosts is significantly skewed towards high mass galaxies, with a significant fraction of events found in passive environments, while core collapse SNe are preferentially hosted in star-forming galaxies with a larger fraction of events found in lower mass galaxies.

\section{Comparison between simulations and the DES photometric sample}
\label{sec:data_vs_sims}

\begin{figure*}
  \centering
\includegraphics[width=0.9\textwidth]{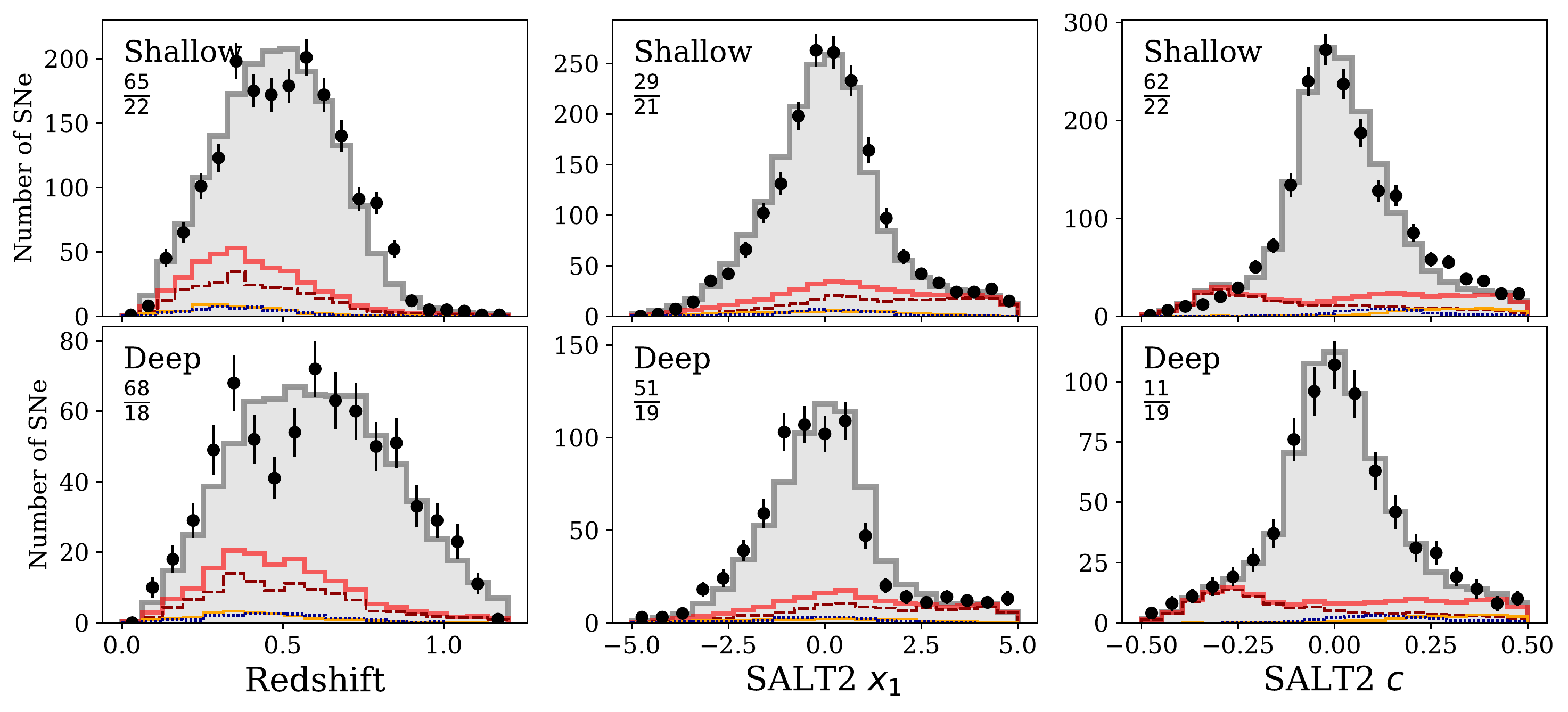}
\includegraphics[width=0.9\textwidth]{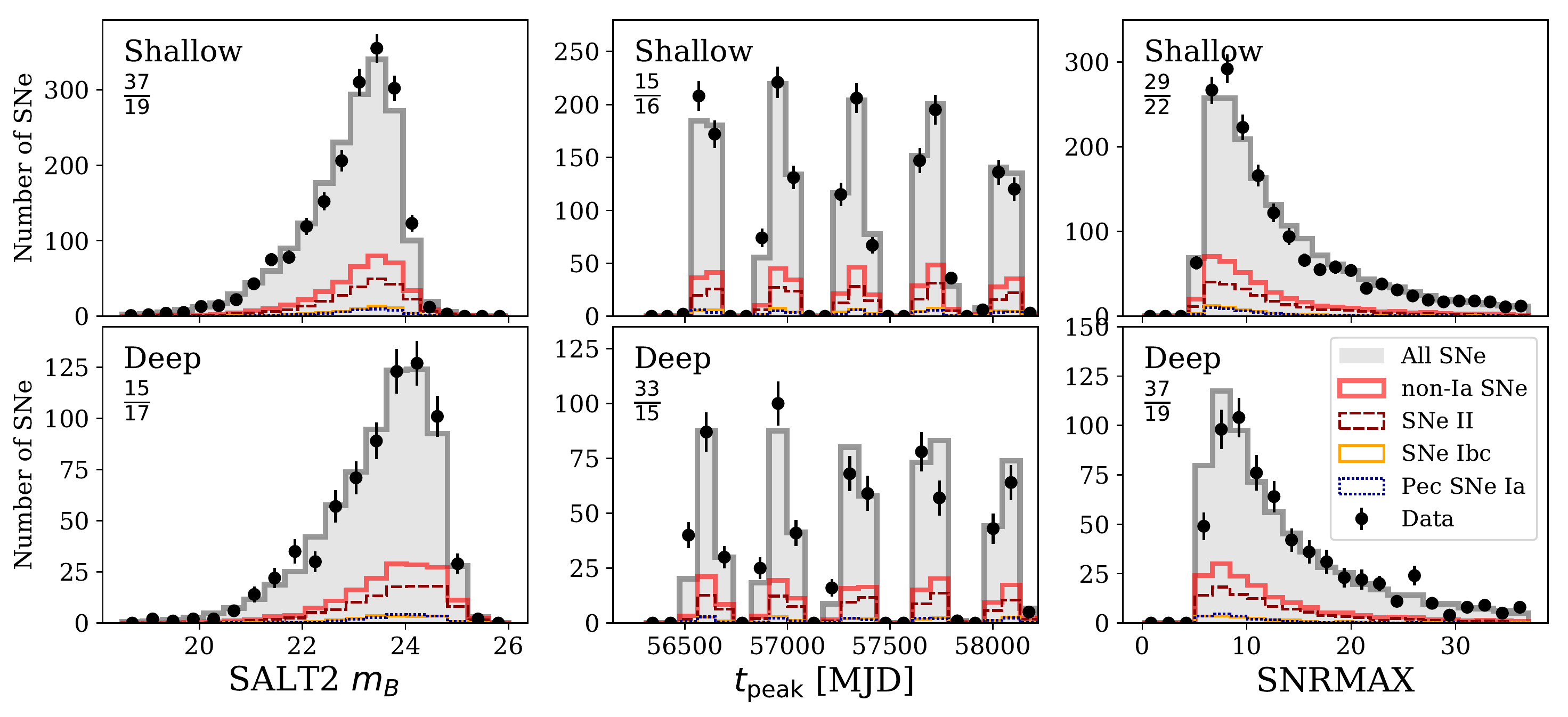}
\includegraphics[width=0.9\textwidth]{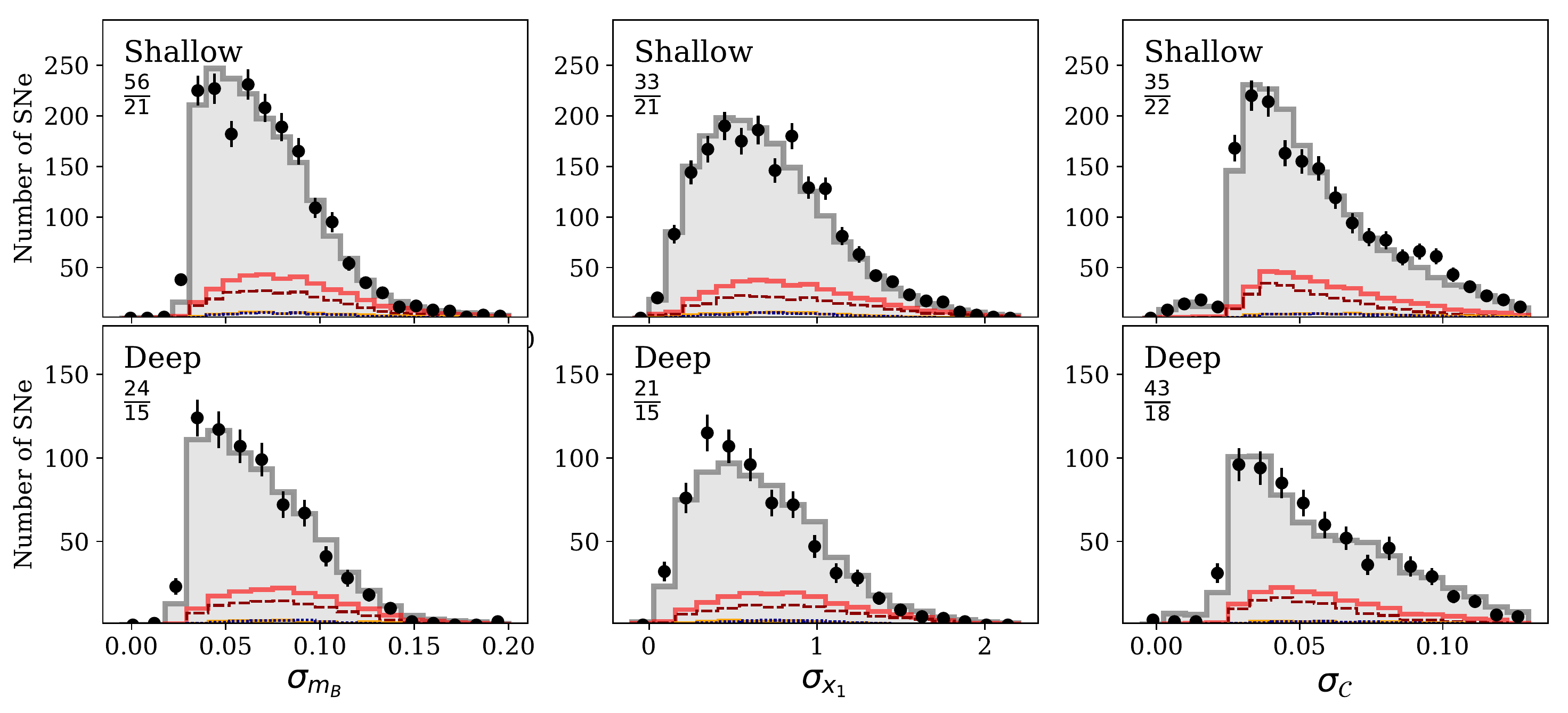}

  \caption{Various comparisons of our \lq Baseline\rq\ simulations and data. The simulations include $\sim$60,000 SNe (25 realisations of the DES photometric SN sample) and the histograms are scaled to match the total number of events in the DES photometric sample. Top panels (from left to right): redshift, SN $x_1$ and SN $c$; central panels: SN $m_B$, MJD of peak brightness, and maximum observed SNR; lower panels (from left to right): uncertainties in the SALT2 fitted parameters $m_B$, $x_1$ and $c$. We compare data (black points), all simulated SNe (SNe~Ia, peculiar SNe~Ia and core collapse SNe combined; grey filled histogram), all non-Ia SNe (solid red line), peculiar SNe~Ia only (SNe 91bg-like and SNe~Iax; dotted line), SNe~Ibc (thin solid orange line) and SNe~II (dashed line). Results are presented for the shallow and deep fields separately. The \rchisq\ is reported in each panel.}
  \label{fig:baseline_comparison_salt2}
\end{figure*}

\begin{figure}
  \centering
\includegraphics[width=0.40\textwidth]{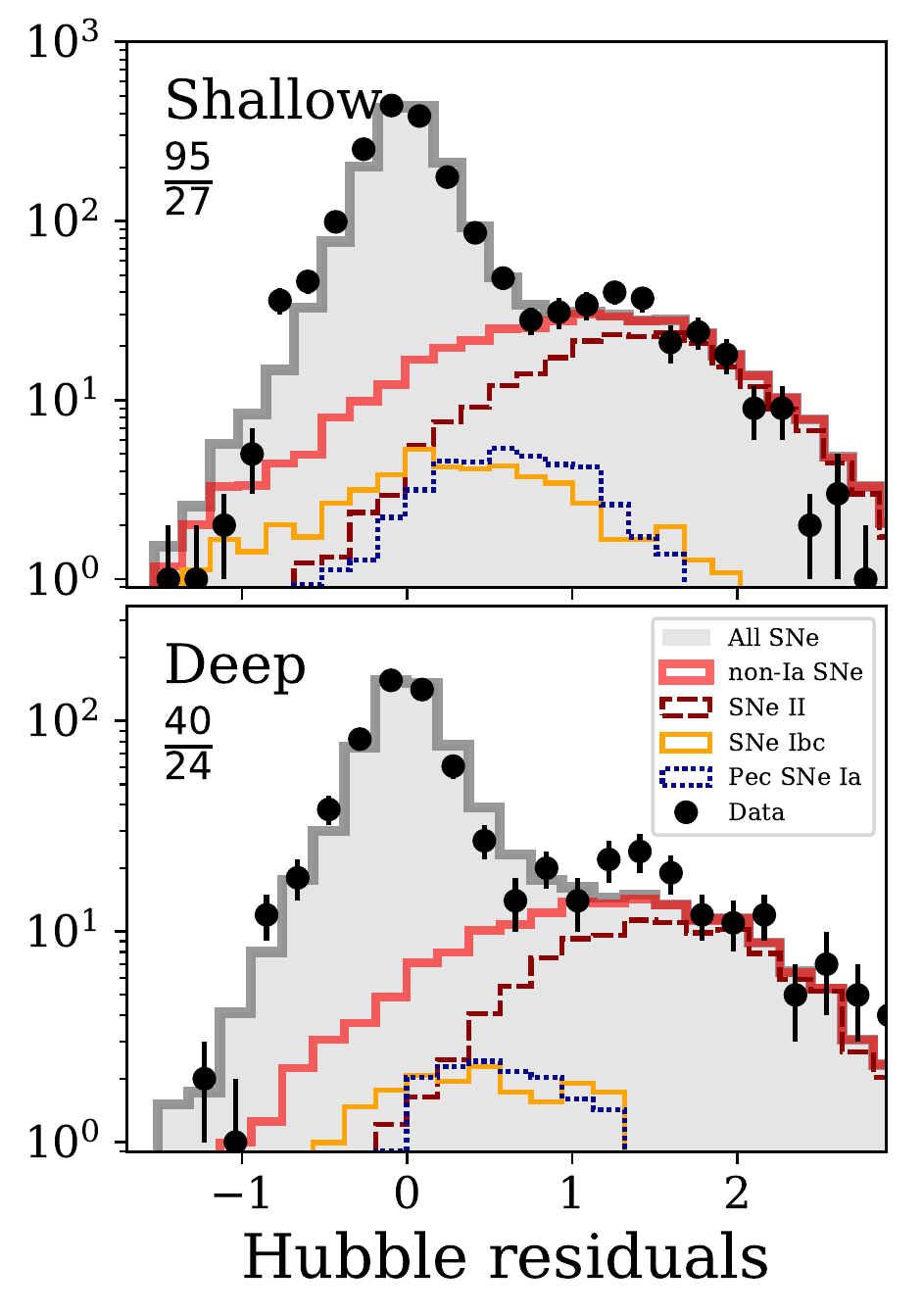}
  \caption{As Fig.~\ref{fig:baseline_comparison_salt2} but for Hubble residuals.}
  \label{fig:baseline_comparison_HR}
\end{figure}

\begin{figure*}
  \centering
\includegraphics[width=1.\textwidth]{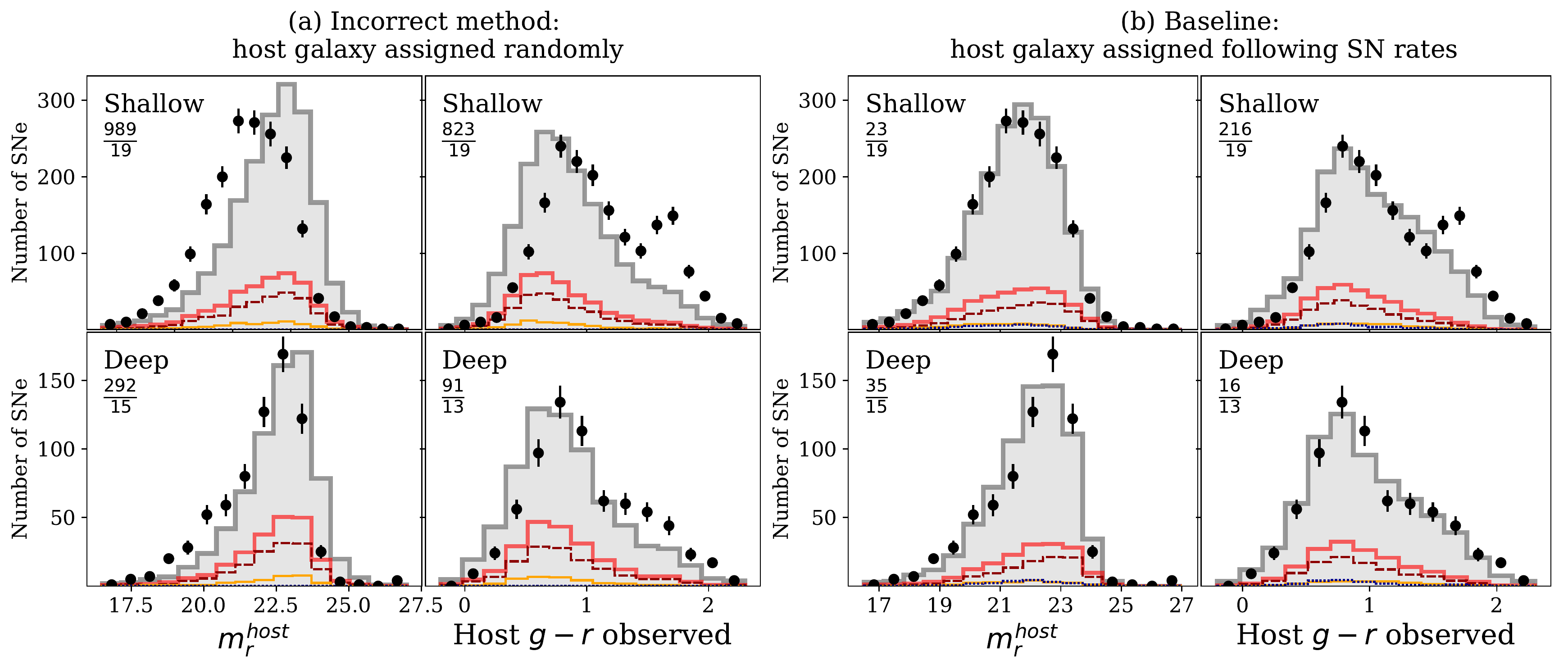}
  \caption{Same as Fig.~\ref{fig:baseline_comparison_salt2}, but for host galaxy $m_r^{\mathrm{host}}$ (left panels) and host galaxy observed $g-r$ colours (right panels). Panel \textbf{(a)} is for an incorrect implementation where host galaxies are assigned randomly to simulated SNe, while panel \textbf{(b)} uses our Baseline simulation.}
  \label{fig:baseline_comparison_host}
\end{figure*}

In Section~\ref{sec:simulations} we presented the Baseline framework of our simulation, the goal of which is to produce a simulation that matches the observed SN populations and properties of the DES photometric SN sample. In this section, we compare our Baseline simulation with the DES photometric SN sample presented in Section~\ref{sec:salt2_data}. This comparison constitutes the core of this paper, and is essential to test the astrophysical assumptions used in our simulations. 

We present the simulation versus data comparisons for distributions of SN redshift, SALT2-fitted SN parameters, and Hubble residuals as described in Section~\ref{sec:salt2_data}. To first order, the Hubble residual distribution of SNe~Ia can be modelled as a symmetric Gaussian, with a mean of zero and a standard deviation equal to the combination of intrinsic scatter of the SN Ia sample and observational noise. Due to the presence of core collapse SN contamination, however, the Hubble residual distribution of a sample of photometrically-classified SNe~Ia will typically have an asymmetrical positive tail \citep{2013ApJ...763...88C, Jones_2017_I}.\footnote{Lensing magnification can also introduce an asymmetrical \emph{negative} tail in the Hubble residual distribution. However, this effect is significantly smaller than the one introduced by core collapse contamination and it is not discussed in this analysis.} Core collapse SNe have, on average, fainter intrinsic brightnesses than SNe~Ia, and are not standardizable using equation~(\ref{eq:tripp}).
 Applying the same equation to an intrinsically fainter SN (like a core collapse SN) leads to an overestimate of the SN distance modulus and thus positive Hubble residual (equation~\ref{eq:hubble_residual}).

In Fig.~\ref{fig:baseline_comparison_salt2} and \ref{fig:baseline_comparison_HR}, we present a comparison between our Baseline simulation and the DES photometric SN sample for the distributions of SALT2 parameters ($m_B$, $x_1$, $c$, $t_{0}$) and their uncertainties, redshift, maximum observed signal-to-noise ratio and Hubble residuals.
In Fig.~\ref{fig:baseline_comparison_host}, the same comparison is presented for $m_{r}^{\mathrm{host}}$ and host galaxy observed $g-r$ colour. We present results for deep and shallow fields separately, using the set of loose SALT2 cuts described in Section~\ref{sec:salt2_data}. We combine 25 realisations of the Baseline simulation (total of 60,000 simulated SNe) and normalise each histogram so that the total number of SNe in the simulation is equal to the total number of observed SNe (for deep and shallow fields separately). We evaluate the level of agreement between data and simulation by calculating the reduced chi-square \rchisq\ (the $\chi^2$ per degree of freedom) as described by \citet[section 3.7.3]{DES_syst}. We report the \rchisq\ in each figure panel.

Qualitatively, the simulation reproduces the DES SN sample well. This is a remarkable result considering the various assumptions that underpin the simulation (e.g., the SN rates, host galaxy properties, SN templates), and considering  the inputs to the simulation have not been tuned to match the data. In detail, in Fig.~\ref{fig:baseline_comparison_salt2} and \ref{fig:baseline_comparison_HR}, we observe:
\begin{itemize}
    \item In the $x_1$ distribution, both data and simulation contain a tail of high-$x_1$ events. This is caused by highly energetic stripped-envelope SNe (SNe Ic, SNe Ic-BL), often characterised by slowly evolving light curves, and by faster-declining SNe~II compared to the general SN II population, but which are still slower that SNe~Ia.
    \item In the $c$ distribution, data and simulations show tails at bluer and redder colours. The bluer tail is caused by SNe~II, similar to hot black bodies at peak and thus with bluer colours than SNe~Ia. The redder tail is mainly due to SNe~Iax and stripped-envelope SNe (see Fig.~\ref{fig:gr_colour_peak_iaPecia} and Fig.~\ref{fig:gr_colour_peak_CC} for a visualisation of where stripped-envelope SNe and SNe~II lie in colour space compared to SNe~Ia).
    \item The distribution of simulated $t_{\mathrm{peak}}$ match the data well, suggesting the time dependency of the spectroscopic redshift efficiency presented in Section~\ref{sec:eff_time} is well modelled.
    \item The faint tail in the Hubble residuals, the clearest feature of the presence of contamination in the data, is also well reproduced. The ratio between the number of SNe with large Hubble residuals ($>0.5$, i.e., likely contaminants) and the number of SNe with small Hubble residuals ($<0.5$, i.e., likely SNe Ia) is 0.20 in data and 0.21 in simulations for the shallow fields. For deep fields, these numbers are 0.34 and 0.30. In photometric SN sample analyses, this is the first time that the contamination observed in the Hubble diagram is explained and almost fully reproduced by a simulation, without the requirement of significant fine tuning of our assumptions and therefore lifting doubts on whether our knowledge of bright core collapse SNe at high redshift present substantial gaps. The only minor discrepancy we observe is that our simulation underestimates the contamination in the deep fields by about 10 per cent. The \rchisq\ is larger than expected from statistical fluctuations, and the excess is mainly driven by the bulk population of SNe~Ia at small Hubble residuals. These discrepancies arise because the Hubble residuals are measured assuming values of the nuisance parameters $\alpha$, $\beta$ and $\mathcal{M}_B$, and assuming a cosmological model.
\end{itemize}

The fact that our simulation reproduces the main features that can be considered signatures of core collapse contamination is promising. Nonetheless, some discrepancies between simulations and observations should be noted.
\begin{itemize}
    \item In the redshift distributions in Fig.~\ref{fig:baseline_comparison_salt2}, we note an underestimate of SN events at high redshift in the shallow fields, and in the deep fields we highlight that the sharp dip observed at redshift z$\sim 0.5$ is not correctly modelled by simulations;
    \item The observed and simulated $x_1$ distributions agree well in the shallow fields but not in the deep fields. Shallow and deep fields probe slightly different redshift ranges and therefore different galaxy populations. SALT2 $x_1$ is known to be correlated with galaxy properties such as galaxy stellar mass, and this discrepancy suggests that our modelling of host mass-$x_1$ correlations and/or the \texttt{HOSTLIB} implemented need to be improved;
    \item Distribution of maximum signal-to-noise ratio shows some discrepancies at lower values, which calls for further improvements in the modelling of flux uncertainties.
\end{itemize}

These discrepancies are unlikely to be solely due to an incorrect modelling of core collapse SNe, as they occur in regions of the parameter space that are primarily dominated by SNe~Ia (e.g., high redshift in the shallow fields, or near zero Hubble residual in the deep fields). 
Further improvements in the modelling of flux uncertainties and selection effects in the DES data may be required, as well as the implementation of a deeper and more complete \texttt{HOSTLIB} that at high redshift will affect the fraction of SNe simulated in faint hosts, i.e., that are unlikely to have a spectroscopic redshift. Further revision of the modelling of SN~Ia intrinsic properties (the intrinsic distributions of $x_1$ and $c$ and the intrinsic scatter) may also be needed. These are all complex aspects of the analysis and we anticipate continued improvements in future analyses.

Finally, Fig.~\ref{fig:baseline_comparison_host}b shows that the observed distribution of 
$m_{r}^{\mathrm{host}}$, is well reproduced by simulations. This agreement suggests that the measurement of spectroscopic efficiency presented in Section~\ref{sec:specz_efficiency} is robust, and that the implemented SN rate models (Section~\ref{sec:sim_hostgalaxies}) adequately describe the data.

The importance of implementing a galaxy-dependent selection in our simulations is demonstrated in Fig.~\ref{fig:baseline_comparison_host}a, the distribution of $m_{r}^{\mathrm{host}}$ from a simulation using the same inputs as the Baseline simulation, but with the exception that host galaxies are assigned randomly (i.e., every galaxy has an equal probability of hosting a SN). Since the \texttt{HOSTLIB} implemented in our simulations is complete to $m_{r}\simeq23.8$\,mag, at redshifts lower than 0.4--0.5 it is dominated by faint and low mass galaxies. As a consequence, a large fraction of SNe is simulated in faint galaxies and are rejected as the OzDES selection function is applied. 
We note that small discrepancies are observed in the distribution of $g-r$ observed colours in the shallow fields, with a fraction of the red galaxies (mostly passive environments, primarily populated by SNe~Ia) missing from simulations. 
This will be further investigated by implementing deeper and higher quality galaxy catalogues in the simulations.

\begin{table}
\centering
    \caption{True fraction of core collapse SNe for different SALT2-based cuts}
    \label{table:contamination_cuts}
    \begin{tabular}{lccc}
\hline
Cut & \multicolumn{3}{c}{Fraction of non-Ia SNe (\%)}\\
\hline
   &  & only this cut & exclude cut \\
   \hline
Loose SALT2 cuts & 22.5 & - & - \\
|$x_1$|<3 & 18.7 & 18.7 & 7.8 \\
|$c$|<0.3 & 13.2 & 16.3 & 10.8 \\
$\sigma_{x_1}<1$ and $\sigma_{t_{\mathrm{peak}}}$<2 & 10.9 & 18.6 & 9.2 \\
Fit prob > 0.01 & 6.6 & 17.3 & 10.9 \\
\end{tabular}
\end{table}

From the Baseline simulation, we can predict the expected core collapse SN contamination in the DES SN Ia sample. Table~\ref{table:contamination_cuts} summarises how this contamination depends on the different SALT2 and light curve cuts that can be applied. For the loose SALT2 cuts, we predict the fraction of non-Ia SNe to be around 22.5 per cent (2.6 per cent arising from peculiar SNe~Ia, 5.7 per cent from SNe~Ibc and 14.2 per cent from SNe~II), and for the \citet{Betoule_2014} SALT2 cuts, the fraction decreases to \baselineCC per cent (1.8 per cent from peculiar SNe~Ia, 1.5 per cent from SNe~Ibc and 3.3 per cent from SNe~II). We highlight that the SALT2 $c$ and fit probability cuts remove the largest fraction of contamination.

SNe~II are the largest source of contamination as they are the most common type of core collapse SN, and the brightest SNe~II are faster declining and therefore photometrically more similar to SNe~Ia than the generally fainter plateauing SNe~II. However, examining the Hubble residual distributions in Fig.~\ref{fig:baseline_comparison_HR} in detail we note that even though SNe~Ibc are not the primary source of contamination, they have on average Hubble residuals closer to zero. In the next section, we discuss how the contamination fraction predicted in the Baseline simulation varies as different assumptions, modelling choices and templates library are used.

\section{Testing alternative core collapse SNe simulations}
\label{sec:sim_cc_vary}

We next analyse how changing the assumptions and modelling choices discussed in Section~\ref{sec:sim_cc_baseline} affects the results of this analysis and in particular the predicted fraction of core collapse SN contamination in the DES sample. We use eight additional core collapse SN simulations generated by adjusting the luminosity functions, the host galaxy dust extinction, the SN colour dispersion, and using different libraries of core collapse SN SED templates. The simulations are summarised in Table~\ref{table:sims}.

\begin{table*}
   \caption{Summary of alternative simulations for core collapse SNe}
   \label{table:sims}
 \centering
\begin{tabular}{lccc}
\hline
Label & Template library & Luminosity functions	&Dust model\\
\hline
   Baseline & \citetalias{Vincenzi_2019} &  revised \citetalias{2011MNRAS.412.1441L}, Gaussian & NA$^*$\\

Skewed LFs &\citetalias{Vincenzi_2019} &  revised \citetalias{2011MNRAS.412.1441L}, skewed Gaussian & NA \\

LFs+Offset &\citetalias{Vincenzi_2019} &  revised \citetalias{2011MNRAS.412.1441L} + offset & NA \\

    LFs $z$-evolving &\citetalias{Vincenzi_2019} &  revised \citetalias{2011MNRAS.412.1441L} + $z$ evolution & NA \\

    Dust (H98)  & dereddened \citetalias{Vincenzi_2019} &  revised \citetalias{2011MNRAS.412.1441L}, Gaussian & \citet{1998ApJ...502..177H}\\

    Dust (R14) & dereddened \citetalias{Vincenzi_2019} &  revised \citetalias{2011MNRAS.412.1441L}, Gaussian & \citet{Rodney_2014}\\

Dust $z$-evolving & dereddened \citetalias{Vincenzi_2019} &  revised \citetalias{2011MNRAS.412.1441L}, Gaussian &  \citet{1998ApJ...502..177H} +$z$ evolution  \\

J17  & \citetalias{Jones_2017_I} &  adjusted LFs from \citetalias{2011MNRAS.412.1441L} & NA\\ 

PLAsTiCC  & PLAsTiCC & PLAsTiCC  & NA \\ 
\hline
    \end{tabular}
    \begin{tablenotes}\footnotesize
        \item $^*$N/A: not applicable -- simulations with core collapse SN templates that are \emph{not} corrected for host dust extinction; additional extinction is not included.
    \end{tablenotes}
\end{table*}

\subsection{Luminosity functions}
\label{sec:sim_cc_vary_LFs}

Luminosity functions, describing the distribution of absolute brightness of the SNe, are a critical element of uncertainty in our analysis. Due to the relative faintness of core collapse SNe and thus the Malmquist biases inherent in SN surveys, luminosity functions are difficult to measure accurately and they depend on whether dust extinction corrections are applied \citep{2011MNRAS.412.1441L, 2014AJ....147..118R}. These corrections are generally uncertain, and it is difficult to disentangle the distribution of intrinsic brightness and the distribution of dust extinction. Currently, published measurements of core collapse SN luminosity functions are based on local SNe (i.e., $<100$\,Mpc). This low-redshift measurement adds further uncertainty as the properties of core collapse SNe may evolve with redshift. 

\begin{figure*}
    \includegraphics[width=0.9\linewidth]{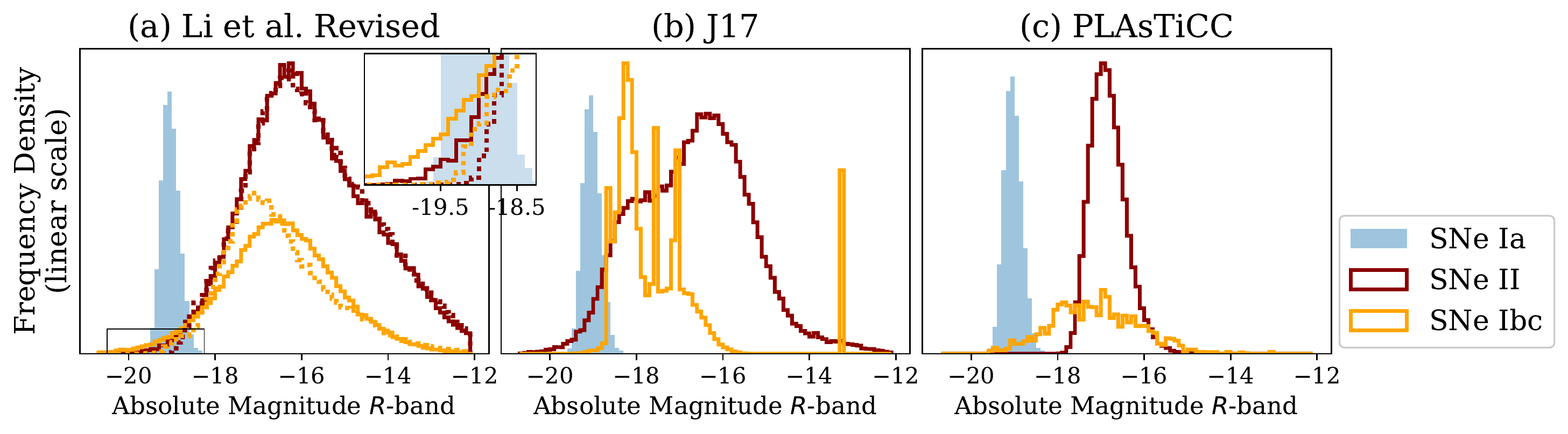}
    \caption{Distributions of simulated $R$-band absolute magnitudes at peak for various types of SNe. This series of panels summarises the different core collapse luminosity functions tested in this work.
    For visualisation purposes we also show the luminosity distribution of SNe~Ia. The relative normalisation between SNe~Ia and core collapse SNe is arbitrary, while the relative rate between stripped-envelope SNe (SNe~Ibc) and hydrogen-rich SNe (SNe~II) is preserved \citep[roughly 0.435, see][]{Shivvers_2017}. In panel \textbf{(a)}, we use the luminosity functions presented by \citetalias{2011MNRAS.412.1441L} and revised by \citetalias{Vincenzi_2019}. We present luminosity distributions derived using both the Gaussian parametrization (Baseline, solid line) and the skewed Gaussian parametrization (Skewed LFs, dotted lines). The analytical forms of the revised \citetalias{2011MNRAS.412.1441L} luminosity functions are summarised in Table~\ref{table:LFs}. The inset in the plot highlights differences in the brightest tail between the two parametrizations. In panel \textbf{(b)} we show luminosity distribution from the \citetalias{Jones_2017_I} core collapse simulations; in panel \textbf{(c)} we show luminosity distributions estimated from simulations generated using the PLAsTICC models (see Section \ref{sec:sim_cc_vary_libraries}).}
    \label{fig:LF_plot}
\end{figure*}

In our analysis, we model luminosity functions based on the volume-limited LOSS sample \citep{Leaman_2011, 2011MNRAS.412.1441L}, taking into account the revised classification published by \citet{Shivvers_2017}. We explore different parametrizations, which we summarise in Table~\ref{table:LFs}:
\begin{itemize}
    \item We assume that the luminosity functions are described by a Gaussian distribution, corresponding to the Baseline simulation presented in Section~\ref{sec:sim_cc_baseline};
    \item We assume that the luminosity functions are described by a skewed Gaussian distribution (\lq Skewed LFs\rq). Table \ref{table:LFs} shows the parameters from skewed luminosity function fits to the revised LOSS sample: mean standard deviation and skeweness. For all sub-types we find a positive skewness, i.e., a larger tail on the fainter side of the luminosity distribution, compatible with the interpretation of dust extinction as the origin.
    \item We apply a redshift-independent offset to the mean of each Gaussian luminosity function measured from the LOSS SN sample (\lq LFs+Offset\rq). The uncertainty on the mean for the LOSS luminosity functions is typically 0.2--0.4\,mag, and therefore adjustments within this range are consistent with the baseline values. However, \citet[][hereafter \citetalias{Jones_2017_I}]{Jones_2017_I} claim that the original LOSS luminosity functions need to be shifted by approximately $-1$\,mag in order to match core collapse SN contamination in the PanSTARRS SN sample. Here we test the choice of an intermediate magnitude shift of $-0.5$\,mag.
    \item We introduce a redshift-dependent drift to the mean of the Gaussian luminosity functions (\lq LF $z$-evolving\rq). This magnitude shift is $\Delta m = -0.5z$\,mag and corresponds to a magnitude offset of $-0.5$\,mag at $z=1$.
\end{itemize}
In addition to the four alternative luminosity functions, we include luminosity functions implemented by \citetalias{Jones_2017_I} and in the PLAsTICC simulations (these simulations are discussed in Section \ref{sec:sim_cc_vary_libraries}), with a total of six luminosity functions tested in this work. These luminosity functions are presented in Fig.~\ref{fig:LF_plot} as distributions of Bessell $R$-band peak absolute magnitudes \citep[for consistency with the luminosity functions presented by][]{2011MNRAS.412.1441L}. The distributions are estimated as follows. We consider the same input luminosity functions and templates designed for the DES core collapse SN simulations tested in this work, and estimate the $R$-band peak absolute magnitudes from a set of 10,000 SN light-curves and examine the distributions. These represent the effective underlying luminosity distributions used in each core collapse SN simulation and allow a direct comparison between different luminosity functions. The distributions presented in the first panel of Fig.~\ref{fig:LF_plot} match the analytical forms presented in Table \ref{table:LFs}. 

\begin{table}
\centering
    \caption{Luminosity functions from \citet{2011MNRAS.412.1441L} with revised classification from \citet{Shivvers_2017}}
    \label{table:LFs}
    \begin{tabular}{lcc}
    SN type  & \multicolumn{2}{c}{Revised LFs from \citet{2011MNRAS.412.1441L}}\\
      &  Gaussian fit $^{a}$ & Skewed gaussian fit $^{b}$ \\
    \hline
    II$^{\dagger}$ &  -15.97(1.31) & -17.51 (2.01,3.18)\\
    IIn &   -17.90(0.95)  & -19.13 (1.53,6.83)\\
    IIb &   -16.69(1.38) & -18.30 (2.03,7.40) \\
    Ic &    -16.75(0.97) & -17.51 (1.24,1.22) \\
    Ib &    -16.07(1.34) & -17.71 (2.11,7.15)\\
    Ic/Ic-pec/Ic-BL  &  -16.79(0.95)  & -17.74 (1.35,2.06) \\
    \hline
    \end{tabular}
    \begin{tablenotes}\footnotesize
        \item $^{a}$ Gaussian fit (mean with standard deviation in parenthesis) of the distributions of $R$-band absolute magnitudes for the bias-corrected LOSS sample. We use the \citet{Shivvers_2017} classifications. Host extinction corrections are not applied.
        \item $^{b}$  Skewed Gaussian fit (mean with standard deviation and skewness in parenthesis) of the distributions of $R$-band absolute magnitudes for the bias-corrected LOSS sample. We use the \citet{Shivvers_2017} classifications. Host extinction corrections are not applied.
        \item $^{\dagger}$ Following the classification scheme introduced by \citet{Anderson_2014} and applied by \citet{Shivvers_2017}, faster declining SNe~II (often referred as SNe~IIL) and slower SNe~II (often referred as SNe~IIP) are combined into a single SN~II class. 
    \end{tablenotes}
\end{table}

\begin{figure*}
    \includegraphics[width=0.9\linewidth]{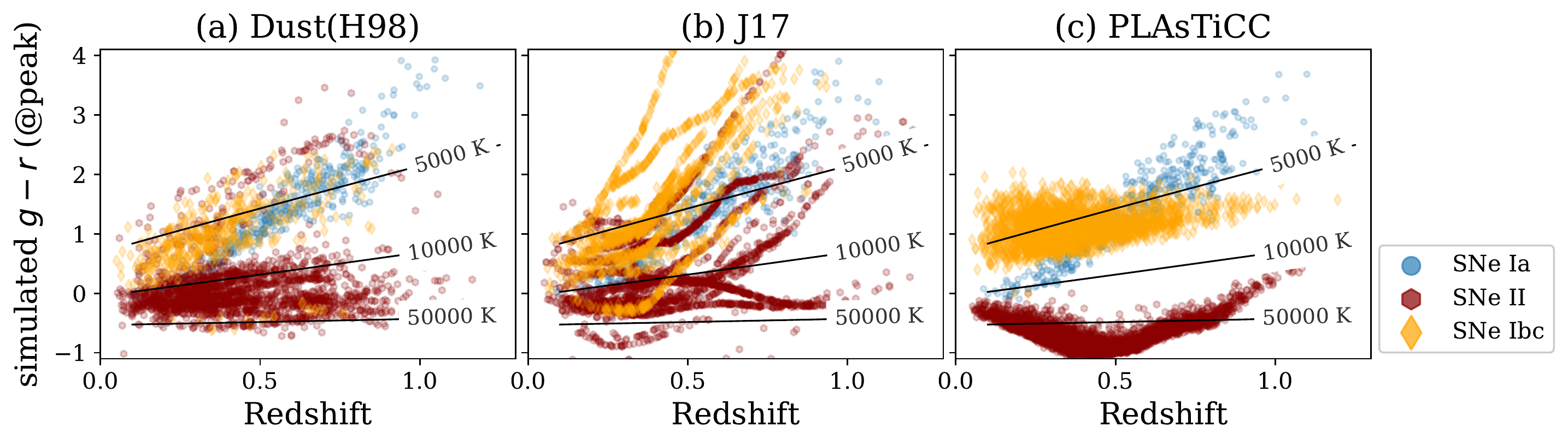}
    \caption{Simulated $g-r$ colour at peak brightness vs redshift for different SN types and templates. SNe~Ia are generated as described in Section~\ref{sec:sim_ia}. Panel \textbf{(a)} is as panel (c) in Fig.~\ref{fig:gr_colour_peak_iaPecia}, but using the \citetalias{Vincenzi_2019} templates and a dust extinction distribution from \citet{1998ApJ...502..177H}. In panel \textbf{(b)}, core collapse SNe are simulated using the \citetalias{Jones_2017_I} set of templates and adjusted luminosity functions; in panel \textbf{(c)} using PLAsTICC models generated using \textsc{mosfit} for stripped-envelope SNe and non-negative matrix factorization for SNe~II (see Section~\ref{sec:sim_cc_vary_libraries}). We also show the $g-r$ colour measured from black body SEDs at temperatures of 5000, 10000 and 50000\,K.}
    \label{fig:gr_colour_peak_CC}
\end{figure*}

\begin{figure}
    \includegraphics[width=0.85\linewidth]{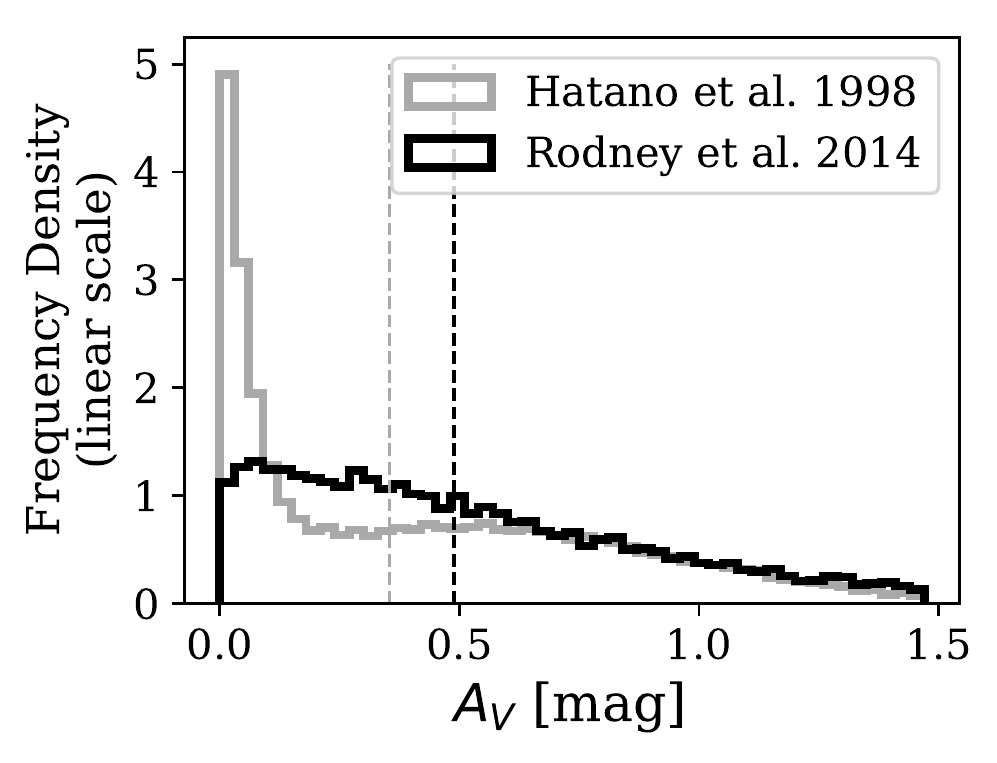}
    \caption{Simulated $A_V$ extinction in Dust(H98) simulation (host dust extinction distribution from \citealp{1998ApJ...502..177H}) and Dust(R14) simulation (host dust extinction distribution from \citealp{Rodney_2014}), see Section~\ref{sec:sim_cc_vary_dust} for more details. Dashed vertical lines show the median simulated $A_V$ for each distribution.}
    \label{fig:dust_plot}
\end{figure}

\subsection{Host galaxy extinction}
\label{sec:sim_cc_vary_dust}

The star-forming hosts of core collapse SNe will typically contain high abundances of gas and dust and thus dust extinction within the host galaxy will be astrophysically important in our simulations. Two sets of \citetalias{Vincenzi_2019} templates are available: one not corrected for host dust extinction  (i.e., implicitly containing some extinction as observed in the SNe) and one corrected for dust extinction \citep[see Appendix A of][for more details]{Vincenzi_2019}.
This allows two implementations of host galaxy extinction and two methods of matching simulated core collapse SNe to luminosity functions. In the first approach, core collapse SN events are simulated with their original host reddening, and the simulated luminosity function is adjusted to match the revised \citetalias{2011MNRAS.412.1441L} luminosity functions. In the second approach, simulated core collapse light-curves are synthesized from the \emph{unreddened} SED models and applying arbitrary extinction models (thus augmenting the diversity, see Fig.~\ref{fig:gr_colour_peak_CC}). The luminosity distribution of the simulated events is matched to the revised \citetalias{2011MNRAS.412.1441L} luminosity functions only \emph{after} the extinction is applied.

We test both approaches and investigate different implementations of host dust extinction:
\begin{itemize}
\item We assume that the host extincted \citetalias{Vincenzi_2019} templates are representative of the core collapse SN population in terms of extinction properties at all redshifts. In other words, we apply no further host extinction. This is our Baseline approach. 
\item We use the set of de-reddened \citetalias{Vincenzi_2019} SEDs and apply the host extinction distribution predicted by \citet{1998ApJ...502..177H} (\lq Dust (H98)\rq). The distribution of $B$-band extinction ($A_B$) presented by \citet{1998ApJ...502..177H} is converted into $A_V$ and fit with the sum of an exponential distribution, $\exp(-A_V/\tau)$, and a normal distribution $\mathcal{N}(\mu,\sigma)$; we find $\tau=0.05$, $\sigma=0.5$ and $\mu=0.45$. Fig.~\ref{fig:dust_plot} shows the resulting distribution of simulated $A_V$. For this model, the median simulated extinction $A_V$ is 0.35\,mag.
\item We use the de-reddened \citetalias{Vincenzi_2019} SEDs and the host extinction distribution used by \citet{Rodney_2014} (\lq Dust (R14)\rq). This distribution is approximated with the same expression adopted for \citet{1998ApJ...502..177H} but assuming $\tau=1.7$, $\sigma=0.6$ and $\mu=0$. Fig.~\ref{fig:dust_plot} shows the resulting distribution of simulated $A_V$. The choice of this distribution results in higher values of extinction, with median simulated $A_V$ of 0.49\,mag. This choice is motivated by the fact that other compilations of core collapse SNe from untargeted surveys (i.e., surveys not primarily based on monitoring bright and typically dust-rich galaxies) seem to have larger mean extinction values \citep{2016MNRAS.458.2973P}.
\item We use the de-reddened \citetalias{Vincenzi_2019} SEDs and the host extinction distribution of \citet{1998ApJ...502..177H}, introducing a redshift dependency in the dust extinction. The dust content of a galaxy correlates with its SFR \citep{Santini_2014}. Since the cosmic star formation increases by $\simeq0.5$\,dex between redshifts 0 and 1 \citep{Madau_2014}, we assume that the median simulated extinction $A_V$ linearly increases by a factor 3 to $z=1$ (\lq Dust $z$-evolving\rq) and apply a shift to the mean of the Gaussian component $\mu$ of $\Delta_{\mu}=0.4z$\,mag.
\end{itemize}
Fig.~\ref{fig:gr_colour_peak_iaPecia}c and Fig.~\ref{fig:gr_colour_peak_CC}a show the simulated $g-r$ colours at peak brightness for different approaches: the Baseline approach, and the approach where the distribution of dust from \citet{1998ApJ...502..177H} is applied on the de-reddened templates (\lq Dust (H98)\rq). In the second case, the diversity of SN events simulated is significantly increased.

\subsection{Comparing different libraries of templates}
\label{sec:sim_cc_vary_libraries}

The most widely used library to date is that of the SN Photometric Classification Challenge \citep[SNPhotCC;][]{2010arXiv1001.5210K,2010PASP..122.1415K}, built from publicly-available composite spectral time series \footnote{\url{https://c3.lbl.gov/nugent/nugent_templates.html}} adjusted to match multi-band photometry for 41 well observed, spectroscopically-confirmed core collapse SNe from various nearby photometric surveys. \citetalias{Jones_2017_I} augmented this library with additional templates of SNe~IIb and 91bg-like SNe~Ia.

In Fig.~\ref{fig:LF_plot}b we show the distribution of $R$-band absolute magnitudes derived from the \citetalias{Jones_2017_I} core collapse SN simulations. \citetalias{Jones_2017_I} simulate SNe~IIb from a set of six SED templates without applying dispersion to the SED brightness, leading to the spikes in the luminosity function, and
assume for SNe Ib a luminosity function with the functional form $\mathcal{N}(-18.26,0.15)$, explaining the brightest peak in the SN Ibc distribution. The bimodality for SNe II is due to SNe~IIP and SNe~IIL being modelled separately, following the rates and luminosity functions originally presented by \citet{2011MNRAS.412.1441L}. We note the \citetalias{Jones_2017_I} templates lack a robust extension into the UV, and therefore at higher redshifts the simulation does not generate $g$-band observations (see Fig.~\ref{fig:gr_colour_peak_CC}b)

\citet{2019PASP..131i4501K} released a new library of core collapse SN templates developed for PLAsTiCC, including two innovative approaches for simulating core collapse SNe. 
For stripped-envelope SNe and SNe~IIn, SED templates have been generated using the Modular Open-Source Fitter for Transients \citep[\textsc{mosfit}; ][]{Guillochon_2018} parametrization and following the theoretical models of \citet{2017ApJ...849...70V} for these two classes of transients. For SNe II, synthetic light curves were built applying dimensionality reduction techniques to a large sample of SN II multi-band light curves.
These techniques enable an order of magnitude increase in the number of SEDs generated (384 templates for SNe~II, 836 for SNe~IIn and stripped-envelope SNe). In Fig.~\ref{fig:LF_plot}c and Fig.~\ref{fig:gr_colour_peak_CC}c we compare luminosity distributions and colour properties of core collapse SNe generated using PLAsTiCC templates with other core collapse SN libraries. We note significant differences both in the distribution of simulated absolute magnitudes and in the colour evolution compared to simulations generated with \citetalias{Vincenzi_2019} and \citetalias{Jones_2017_I} templates.

\subsection{Analysis of Hubble residuals distributions}
\label{sec:HR_analysis}

\begin{table*}
   \caption{True fraction of contamination (averaged over 25 realisations).}
   \label{table:data_sims_numbers}
 \centering
\begin{tabular}{lclc|clc}
 & \multicolumn{3}{c|}{Loose SALT2 cuts} &  \multicolumn{3}{c}{SALT2 cuts following \citet{Betoule_2014}} \\
 \hline
  &  \multicolumn{1}{c}{Non-Ia fraction} &
  \multicolumn{1}{c}{Fraction of} &
   \multicolumn{1}{c|}{Non-Ia fraction in} &
   \multicolumn{1}{c}{Non-Ia fraction} &
   \multicolumn{1}{c}{Fraction of } &
   \multicolumn{1}{c}{Non-Ia fraction in} \\

  &  \multicolumn{1}{c}{(\%)} &
  \multicolumn{1}{c}{91bg, Iax, Ibc, II (\%)} &
   \multicolumn{1}{c|}{Shallow\&Deep (\%)} &
   \multicolumn{1}{c}{(\%)} &
   \multicolumn{1}{c}{Iax, Ibc, II $^{\dagger}$ (\%)} &
   \multicolumn{1}{c}{Shallow\&Deep (\%)} \\

\hline
Baseline & 22.5 & 0.1, 2.5, 5.7, 14.2 & 21.6, 24.5 & 6.6 & 1.8, 1.5, 3.3 & 6.3, 7.2 \\
Skewed LFs & 20.4 & 0.1, 2.6, 4.4, 13.2 & 19.5, 22.5 & 6.0 & 1.8, 1.2, 3.0 & 5.8, 6.5 \\
LFs z-evolving & 27.5 & 0.1, 2.4, 7.1, 17.8 & 26.4, 30.0 & 8.0 & 1.7, 2.0, 4.3 & 7.7, 8.8 \\
LFs+Offset & 31.7 & 0.1, 2.2, 8.6, 20.7 & 30.8, 33.6 & 9.3 & 1.7, 2.7, 4.9 & 9.0, 10.0 \\
Dust(H98) & 22.0 & 0.1, 2.6, 6.1, 13.2 & 21.1, 24.1 & 6.9 & 1.8, 1.9, 3.2 & 6.6, 7.5 \\
Dust(R14) & 21.6 & 0.1, 2.6, 5.6, 13.3 & 20.8, 23.6 & 6.7 & 1.8, 1.6, 3.4 & 6.3, 7.8 \\
Dust z-evolving & 18.6 & 0.1, 2.7, 4.9, 10.9 & 17.8, 20.4 & 5.8 & 1.8, 1.3, 2.7 & 5.6, 6.3 \\
J17 (PanSTARRS) & 29.1 & 0.1, 2.3, 12.0, 14.7 & 27.9, 31.9 & 7.3 & 1.8, 3.1, 2.5 & 6.7, 8.9 \\
PLAsTiCC & 24.6 & 0.1, 2.5, 7.3, 14.3 & 23.1, 27.9 & 5.6 & 1.8, 1.7, 2.0 & 5.2, 6.5 \\

\hline
\end{tabular}
    \begin{tablenotes}\footnotesize
     \item $^{\dagger}$ After SALT2-based cuts following \citet{Betoule_2014} are applied, the predicted fraction of 91bg-like SNe~Ia is less than 0.1 per cent.
    \end{tablenotes}
\end{table*}

\begin{table}
   \caption{\rchisq\ between observed and simulated events for different Hubble residual ranges.}
   \label{table:chi2}
 \centering
\begin{tabular}{lcccc}
\hline
 & \multicolumn{2}{c}{Loose SALT2 cuts} &  \multicolumn{2}{c}{\citet{Betoule_2014}} \\
 & \multicolumn{2}{c}{ } &  \multicolumn{2}{c}{SALT2 cuts } \\
 \hline 
 & \multicolumn{2}{c}{\rchisq} & \multicolumn{2}{c}{\rchisq} \\
 & \multicolumn{1}{c}{HR$<0.5$} &  \multicolumn{1}{c}{HR$>0.5$}  & \multicolumn{1}{c}{HR$<0.5$} &  \multicolumn{1}{c}{HR$>0.5$} \\

\hline
\emph{SNe~Ia only}  & 5.3 & 60.3  & 3.9 & 17.2 \\
\emph{Peculiar~Ia only}$^{\dagger}$& 5.0 & 29.3  & 3.9 & 5.9 \\
Baseline  & 4.2 & 1.9  & 3.7 & 0.8 \\
Skewed LFs  & 4.9 & 1.5  & 3.8 & 0.7 \\
LFs $z$-evolving  & 3.9 & 1.8  & 3.6 & 1.0 \\
LFs+Offset  & 4.1 & 1.8  & 3.5 & 0.9 \\
Dust (H98)  & 4.0 & 1.7  & 3.7 & 0.9 \\
Dust (R14)  & 4.0 & 2.4  & 3.6 & 1.2 \\
Dust $z$-evolving  & 4.2 & 2.0  & 3.7 & 1.1 \\
J17 (PanSTARRS)  & 6.7 & 3.0  & 4.2 & 1.2 \\
PLAsTiCC  & 5.5 & 10.3  & 4.0 & 1.5 \\

\hline
\end{tabular}
   \begin{tablenotes}\footnotesize
        \item $^{\dagger}$ Simulation generated including only SNe Ia and peculiar SNe Ia, SNe~Iax and 91bg-like SNe Ia.
   \end{tablenotes}
\end{table}

\begin{figure*}
  \centering
  \includegraphics[width=0.99\linewidth]{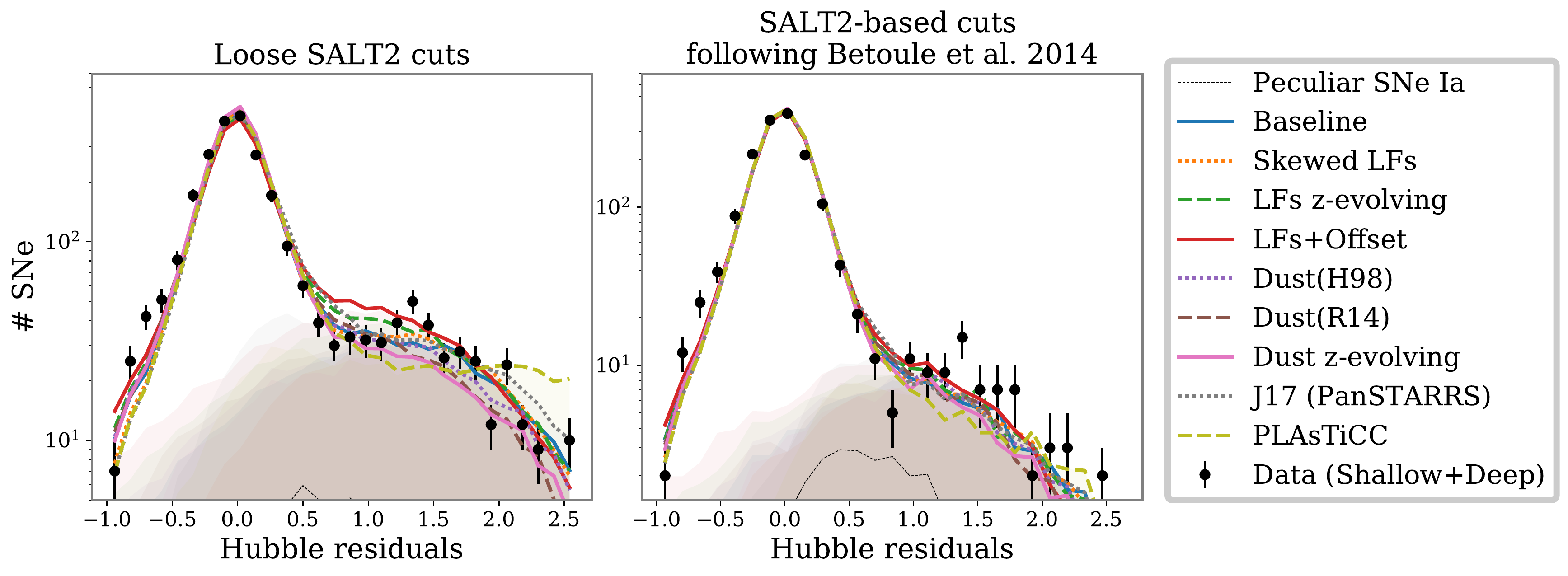}
  \caption{Distributions of observed and simulated Hubble residuals for the full range of simulations. Distributions are presented for the data (shallow and deep field combined, black symbols) and for the 9 simulations summarised in Table~\ref{table:sims} (see Section~\ref{sec:sim_cc_vary}): both SNe~Ia and core collapse SNe are combined in the solid lines, and simulated core collapse SNe only are shown as shaded areas. 
    \emph{Left:} Sample selected applying loose SALT2-based cuts ($x_1 \in [-4.9,4.9]$ and SALT2 $c \in [-0.49,0.49]$). The fraction of core collapse SNe for each simulation is reported in Table~\ref{table:data_sims_numbers} and is approximately a fourth of the sample. \emph{Right:} SALT2-based cuts from \citet{Betoule_2014} are applied. The fraction of core collapse SNe in the simulations is reported in Table~\ref{table:data_sims_numbers}.
    }
  \label{fig:datavssim}
\end{figure*}

In Fig.~\ref{fig:datavssim}, we present the simulated and observed Hubble residuals (equation~\ref{eq:hubble_residual}) for each simulation (Table~\ref{table:sims}) and for the different SALT2 cuts (Section~\ref{sec:salt2_data}). Table~\ref{table:data_sims_numbers} presents the predicted fraction of contamination from 91bg-like, SNe~Iax, SNe~Ibc and SNe~II, and the total contamination, for shallow and deep fields separately. Finally, Table~\ref{table:chi2} presents the \rchisq\ of Hubble residual distributions. \rchisq\ are estimated both for Hubble residuals $<0.5$ (the \lq SN~Ia dominated\rq\ region) and $>0.5$ (the \lq core collapse SN dominated\rq\ region).

Generally, the agreement is good. As noted for the Baseline simulation, the largest discrepancies are found at zero and negative Hubble residuals where the contamination is small (Fig.~\ref{fig:datavssim}), and this drives the large value of \rchisq\ (Table \ref{table:chi2}).
When loose SALT2 cuts are applied, more significant discrepancies are found in the core collapse SN simulations where the luminosity functions are artificially brightened (\lq LFs $z$-evolving\rq\ and \lq LFs+Offset\rq). These simulations overestimate the number of SNe with Hubble residuals $>0.5$ by approximately 20--25 per cent, disfavouring such adjustments. Simulations where larger host extinctions are applied (\lq Dust (R14)\rq\ and \lq Dust $z$-evolving\rq) underestimate the number of SNe with Hubble residuals $<0.5$ by 10 per cent.

When the cuts from \citet{Betoule_2014} are applied, the simulations accurately predict the number of events with large Hubble residuals (HR$>0.5$), with \rchisq\ values between 0.7 to 1.2 (Table \ref{table:data_sims_numbers}). 
The large discrepancies observed when applying only loose SALT2-based cuts in simulations (LFs+Offset and LFs $z$-evolving) appear to be partially resolved when tighter SALT2 cuts are applied. This suggests that understanding how SALT2-based cuts affect core collapse contamination is an important aspect in this type of analysis. 

The \lq Skewed LFs\rq\ simulation predicts one of the lowest values of core collapse SN contamination. As shown in Fig.~\ref{fig:LF_plot}, a skewed Gaussian parametrization of the luminosity functions produces less bright events compared to a Gaussian parametrization of the luminosity functions. This shows that simulation of core collapse SNe is sensitive to how the brighter tails of the luminosity functions are modelled. Finally, we note that the PLAsTiCC simulation shows poor agreement with the data, both before and after SALT2-base cuts.

Overall, the range of contamination predicted by our simulations is small, with a minimum of \minCC per cent of contamination predicted from the \lq Dust $z$-evolving LFs\rq\ simulation (excluding PLAsTiCC simulation) to a maximum of \maxCC per cent contamination in the \lq LFs+Offset\rq\ simulation. The average contamination among the different tested scenarios is \avgCC per cent, and the r.m.s is \rmsCC per cent. We note that this is the contamination expected in the photometric DES SN sample prior to the application of any photometric classification algorithm. After photometric classification, the typical contamination expected is likely to decrease substantially \citep[][Vincenzi et al. in prep]{2020MNRAS.491.4277M}.

\section{Summary and future work}
\label{sec:summary}

We have presented a set of simulations designed to reproduce the DES photometric SN sample. The DES photometric SN sample includes more than 2500 SNe with high-quality multi-band photometry and spectroscopic redshifts from the identified SN host galaxies. It is the largest sample of photometrically identified SNe~Ia to date and in this work we described how the sample has been collected and what are the relevant cuts we applied to filter non-SN transients. 

Our focus in this paper has been to model and reproduce with simulations the population of contaminants observed in the DES photometric sample, where we define as contaminants transients that are photometrically similar to SNe~Ia but are not standardisable candles, i.e., peculiar SNe~Ia and core collapse SNe. The simulations presented in this work are a significant improvement compared to previously published mock catalogues of photometric SN samples. The principle advances are:
\begin{enumerate}
    \item We use core collapse SNe that are synthesised from high-quality templates. We explore different methods for implementing host galaxy dust extinction and different luminosity functions, and we demonstrate that the diversity and quality of the simulated core collapse SN light-curves are significantly improved;
    \item We use a host-galaxy spectroscopic redshift efficiency that is modelled as a three-dimensional function of host galaxy brightness, observed colours and year of SN discovery. This efficiency function has been measured by analysing the sample of galaxies that hosted DES SN candidates and comparing those for which a spectroscopic redshift was obtained and those for which it was not;
    \item We simulate SN host galaxies using published SN rates and their dependency on host galaxy properties. This ensures that each sub-type of transient is associated with a physically meaningful population of galaxies. This, combined with our measured efficiency function, enable us to accurately model selection effects for every type of transient, every type of galaxy, and every redshift range.
\end{enumerate}
The ultimate test to verify if our simulations are realistic and physically accurate is to compare the simulated samples with the real data. We find excellent agreement between our simulations and the DES SN sample, both when loose and cosmology-like SALT2-based cuts are applied.
From our baseline simulation, we predict the fraction of core collapse SN contamination in the DES SN sample to be \baselineCC per cent after applying SALT2-based cuts similar to those in the cosmological analysis from \citet{Betoule_2014}. 

We additionally explore alternative template libraries, luminosity functions and host galaxy dust extinction models. We consider nine core collapse SN scenarios, designed to span a wide range of modelling choices. We analyse this set of simulations and find that the majority reproduce observed contamination well (with measured \rchisq\ between 0.8 and 1.4 for large Hubble residuals) and that the predicted core collapse contamination varies between \minCC and \maxCC per cent, with an average of \avgCC and an r.m.s. of \rmsCC. This suggests that, although our knowledge of the global properties of core collapse SNe remains incomplete, core collapse SN contamination in the DES photometric SN sample can be well constrained.

While the agreement between data and simulations is already good, some discrepancies remain and we anticipate improvements from future analyses. Different ways of increasing the depth of the galaxy library implemented in our simulations will be explored, either using observations \citep[i.e. deep coadds published by][]{DES_deepstacks} or simulations \citep{skypy_collaboration_2020_3755531}.
Additionally, the modelling of SN~Ia intrinsic properties and contribution of the host galaxy to the observational noise will be studied.  

This work lays the foundation for several analyses central to the cosmological analysis of the DES photometric SN survey. 
The cosmological constraints obtained will depend more on our ability to validate the true contamination rather than obtaining the smallest prediction for that contamination.
In future papers, we will use the simulations presented here to train and test the photometric classifiers that will be implemented in the cosmological analysis of the DES SN sample. We will also measure systematic uncertainties and potential biases in cosmological measurements due to core collapse SN contamination. Finally, the methods and techniques used in this work constitute a powerful tool to predict core collapse SN contamination in future cosmological SN Ia samples and can be applied to simulate SNe in time-domain surveys like the 10-year Legacy Survey of Space and Time \citep[LSST]{2019ApJ...873..111I} and surveys with the Nancy Grace Roman Space Telescope \citep{2018ApJ...867...23H}.

\section*{Acknowledgements}

This work was supported by the Science and Technology Facilities Council [grant number ST/P006760/1] through the DISCnet Centre for Doctoral Training. MS acknowledges support from EU/FP7-ERC grant 615929, and PW acknowledges support from STFC grant ST/R000506/1.
L.G. was funded by the European Union's Horizon 2020 research and innovation programme under the Marie Sk\l{}odowska-Curie grant agreement No. 839090. This work has been partially supported by the Spanish grant PGC2018-095317-B-C21 within the European Funds for Regional Development (FEDER).

This paper has gone through internal review by the DES collaboration.
Funding for the DES Projects has been provided by the U.S. Department of Energy, the U.S. National Science Foundation, the Ministry of Science and Education of Spain, 
the Science and Technology Facilities Council of the United Kingdom, the Higher Education Funding Council for England, the National Center for Supercomputing 
Applications at the University of Illinois at Urbana-Champaign, the Kavli Institute of Cosmological Physics at the University of Chicago, 
the Center for Cosmology and Astro-Particle Physics at the Ohio State University,
the Mitchell Institute for Fundamental Physics and Astronomy at Texas A\&M University, Financiadora de Estudos e Projetos, 
Funda{\c c}{\~a}o Carlos Chagas Filho de Amparo {\`a} Pesquisa do Estado do Rio de Janeiro, Conselho Nacional de Desenvolvimento Cient{\'i}fico e Tecnol{\'o}gico and 
the Minist{\'e}rio da Ci{\^e}ncia, Tecnologia e Inova{\c c}{\~a}o, the Deutsche Forschungsgemeinschaft and the Collaborating Institutions in the Dark Energy Survey. 

The Collaborating Institutions are Argonne National Laboratory, the University of California at Santa Cruz, the University of Cambridge, Centro de Investigaciones Energ{\'e}ticas, 
Medioambientales y Tecnol{\'o}gicas-Madrid, the University of Chicago, University College London, the DES-Brazil Consortium, the University of Edinburgh, 
the Eidgen{\"o}ssische Technische Hochschule (ETH) Z{\"u}rich, 
Fermi National Accelerator Laboratory, the University of Illinois at Urbana-Champaign, the Institut de Ci{\`e}ncies de l'Espai (IEEC/CSIC), 
the Institut de F{\'i}sica d'Altes Energies, Lawrence Berkeley National Laboratory, the Ludwig-Maximilians Universit{\"a}t M{\"u}nchen and the associated Excellence Cluster Universe, 
the University of Michigan, NFS's NOIRLab, the University of Nottingham, The Ohio State University, the University of Pennsylvania, the University of Portsmouth, 
SLAC National Accelerator Laboratory, Stanford University, the University of Sussex, Texas A\&M University, and the OzDES Membership Consortium.

Based in part on observations at Cerro Tololo Inter-American Observatory at NSF's NOIRLab (NOIRLab Prop. ID 2012B-0001; PI: J. Frieman), which is managed by the Association of Universities for Research in Astronomy (AURA) under a cooperative agreement with the National Science Foundation.

The DES data management system is supported by the National Science Foundation under Grant Numbers AST-1138766 and AST-1536171.
The DES participants from Spanish institutions are partially supported by MICINN under grants ESP2017-89838, PGC2018-094773, PGC2018-102021, SEV-2016-0588, SEV-2016-0597, and MDM-2015-0509, some of which include ERDF funds from the European Union. IFAE is partially funded by the CERCA program of the Generalitat de Catalunya.
Research leading to these results has received funding from the European Research
Council under the European Union's Seventh Framework Program (FP7/2007-2013) including ERC grant agreements 240672, 291329, and 306478.
We  acknowledge support from the Brazilian Instituto Nacional de Ci\^encia
e Tecnologia (INCT) do e-Universo (CNPq grant 465376/2014-2).

This manuscript has been authored by Fermi Research Alliance, LLC under Contract No. DE-AC02-07CH11359 with the U.S. Department of Energy, Office of Science, Office of High Energy Physics.

Finally, this work was based in part on data acquired at the Anglo-Australian Telescope, under program A/2013B/012. We acknowledge the traditional owners of the land on which the AAT stands, the Gamilaraay people, and pay our respects to elders past and present.

\section*{Data availability statement}
Input and configuration files needed to reproduce the simulations using the software \snana are made available at \url{https://github.com/maria-vincenzi/DES_CC_simulations}. Data relative to the DES photometric sample used in Figures \ref{fig:baseline_comparison_salt2}, \ref{fig:baseline_comparison_HR}, \ref{fig:baseline_comparison_host} and \ref{fig:datavssim} are also made available (counts per each bin for all distributions presented in the Figures).

\appendix
\section{AUTHOR AFFILIATIONS}
\label{aff}
$^{1}$ Institute of Cosmology and Gravitation, University of Portsmouth, Portsmouth, PO1 3FX, UK\\
$^{2}$ School of Physics and Astronomy, University of Southampton,  Southampton, SO17 1BJ, UK\\
$^{3}$ Department of Astrophysics, American Museum of Natural History, New York, NY, USA\\
$^{4}$ NASA Einstein Fellow\\
$^{5}$ Center for Astrophysics, Harvard \& Smithsonian, 60 Garden Street, Cambridge, MA 02138, USA\\
$^{6}$ School of Mathematics and Physics, University of Queensland,  Brisbane, QLD 4072, Australia\\
$^{7}$ PITT PACC, Department of Physics and Astronomy, University of Pittsburgh, Pittsburgh, PA 15260, USA\\
$^{8}$ Space Telescope Science Institute, Baltimore, MD, USA\\
$^{9}$ Department of Astronomy and Astrophysics, University of Chicago, Chicago, IL 60637, USA\\
$^{10}$ Kavli Institute for Cosmological Physics, University of Chicago, Chicago, IL 60637, USA\\
$^{11}$ Argonne National Laboratory, 9700 South Cass Avenue, Lemont, IL 60439, USA\\
$^{12}$ Centre for Gravitational Astrophysics, College of Science, The Australian National University, ACT 2601, Australia\\
$^{13}$ The Research School of Astronomy and Astrophysics, Australian National University, ACT 2601, Australia\\
$^{14}$ Universit\'e Clermont Auvergne, CNRS/IN2P3, LPC, F-63000 Clermont-Ferrand, France\\
$^{15}$ Department of Physics and Astronomy, University of Pennsylvania, Philadelphia, PA 19104, USA\\
$^{16}$ Department of Physics, Duke University Durham, NC 27708, USA\\
$^{17}$ Centro de Investigaciones Energ\'eticas, Medioambientales y Tecnol\'ogicas (CIEMAT), Madrid, Spain\\
$^{18}$ Sydney Institute for Astronomy, School of Physics, A28, The University of Sydney, NSW 2006, Australia\\
$^{19}$ Departamento de F\'isica Matem\'atica, Instituto de F\'isica, Universidade de S\~ao Paulo, CP 66318, S\~ao Paulo, SP, 05314-970, Brazil\\
$^{20}$ Laborat\'orio Interinstitucional de e-Astronomia - LIneA, Rua Gal. Jos\'e Cristino 77, Rio de Janeiro, RJ - 20921-400, Brazil\\
$^{21}$ Fermi National Accelerator Laboratory, P. O. Box 500, Batavia, IL 60510, USA\\
$^{22}$ Instituto de Fisica Teorica UAM/CSIC, Universidad Autonoma de Madrid, 28049 Madrid, Spain\\
$^{23}$ CNRS, UMR 7095, Institut d'Astrophysique de Paris, F-75014, Paris, France\\
$^{24}$ Sorbonne Universit\'es, UPMC Univ Paris 06, UMR 7095, Institut d'Astrophysique de Paris, F-75014, Paris, France\\
$^{25}$ Department of Physics \& Astronomy, University College London, Gower Street, London, WC1E 6BT, UK\\
$^{26}$ Kavli Institute for Particle Astrophysics \& Cosmology, P. O. Box 2450, Stanford University, Stanford, CA 94305, USA\\
$^{27}$ SLAC National Accelerator Laboratory, Menlo Park, CA 94025, USA\\
$^{28}$ Instituto de Astrofisica de Canarias, E-38205 La Laguna, Tenerife, Spain\\
$^{29}$ Universidad de La Laguna, Dpto. Astrofísica, E-38206 La Laguna, Tenerife, Spain\\
$^{30}$ Department of Astronomy, University of Illinois at Urbana-Champaign, 1002 W. Green Street, Urbana, IL 61801, USA\\
$^{31}$ National Center for Supercomputing Applications, 1205 West Clark St., Urbana, IL 61801, USA\\
$^{32}$ Institut de F\'{\i}sica d'Altes Energies (IFAE), The Barcelona Institute of Science and Technology, Campus UAB, 08193 Bellaterra (Barcelona) Spain\\
$^{33}$ Institut d'Estudis Espacials de Catalunya (IEEC), 08034 Barcelona, Spain\\
$^{34}$ Institute of Space Sciences (ICE, CSIC),  Campus UAB, Carrer de Can Magrans, s/n,  08193 Barcelona, Spain\\
$^{35}$ Center for Cosmology and Astro-Particle Physics, The Ohio State University, Columbus, OH 43210, USA\\
$^{36}$ INAF-Osservatorio Astronomico di Trieste, via G. B. Tiepolo 11, I-34143 Trieste, Italy\\
$^{37}$ Institute for Fundamental Physics of the Universe, Via Beirut 2, 34014 Trieste, Italy\\
$^{38}$ Observat\'orio Nacional, Rua Gal. Jos\'e Cristino 77, Rio de Janeiro, RJ - 20921-400, Brazil\\
$^{39}$ Department of Physics, University of Michigan, Ann Arbor, MI 48109, USA\\
$^{40}$ Department of Physics, IIT Hyderabad, Kandi, Telangana 502285, India\\
$^{41}$ Santa Cruz Institute for Particle Physics, Santa Cruz, CA 95064, USA\\
$^{42}$ Institute of Theoretical Astrophysics, University of Oslo. P.O. Box 1029 Blindern, NO-0315 Oslo, Norway\\
$^{43}$ Department of Astronomy, University of Michigan, Ann Arbor, MI 48109, USA\\
$^{44}$ Department of Physics, Stanford University, 382 Via Pueblo Mall, Stanford, CA 94305, USA\\
$^{45}$ Department of Physics, The Ohio State University, Columbus, OH 43210, USA\\
$^{46}$ Faculty of Physics, Ludwig-Maximilians-Universit\"at, Scheinerstr. 1, 81679 Munich, Germany\\
$^{47}$ Max Planck Institute for Extraterrestrial Physics, Giessenbachstrasse, 85748 Garching, Germany\\
$^{48}$ Universit\"ats-Sternwarte, Fakult\"at f\"ur Physik, Ludwig-Maximilians Universit\"at M\"unchen, Scheinerstr. 1, 81679 M\"unchen, Germany\\
$^{49}$ Center for Astrophysics $\vert$ Harvard \& Smithsonian, 60 Garden Street, Cambridge, MA 02138, USA\\
$^{50}$ Australian Astronomical Optics, Macquarie University, North Ryde, NSW 2113, Australia\\
$^{51}$ Lowell Observatory, 1400 Mars Hill Rd, Flagstaff, AZ 86001, USA\\
$^{52}$ Department of Astronomy, The Ohio State University, Columbus, OH 43210, USA\\
$^{53}$ Radcliffe Institute for Advanced Study, Harvard University, Cambridge, MA 02138\\
$^{54}$ Instituci\'o Catalana de Recerca i Estudis Avan\c{c}ats, E-08010 Barcelona, Spain\\
$^{55}$ Physics Department, 2320 Chamberlin Hall, University of Wisconsin-Madison, 1150 University Avenue Madison, WI  53706-1390\\
$^{56}$ Institute of Astronomy, University of Cambridge, Madingley Road, Cambridge CB3 0HA, UK\\
$^{57}$ Department of Astrophysical Sciences, Princeton University, Peyton Hall, Princeton, NJ 08544, USA\\
$^{58}$ Department of Physics and Astronomy, Pevensey Building, University of Sussex, Brighton, BN1 9QH, UK\\
$^{59}$ Computer Science and Mathematics Division, Oak Ridge National Laboratory, Oak Ridge, TN 37831\\
$^{60}$ Cerro Tololo Inter-American Observatory, NSF's National Optical-Infrared Astronomy Research Laboratory, Casilla 603, La Serena, Chile.

\bibliographystyle{mnras}
\bibliography{references.bib}

\bsp	
\label{lastpage}
\end{document}